\documentclass[aps,preprint]{revtex4}
\usepackage{amsmath}
\usepackage{graphicx,color}
\usepackage{booktabs}
\usepackage{datetime}
\definecolor{navyblue}{rgb}{0.0,0.0,1}
\usepackage{xcolor}
\usepackage{epstopdf}
\usepackage{wrapfig}
\usepackage{ragged2e}
\usepackage{epsfig}
\usepackage{slashed}
\usepackage{caption}

\usepackage{hyperref}
\hypersetup{
colorlinks,
citecolor=black,
filecolor=black,
linkcolor=black,
urlcolor=blue}

\begin{document}
\title{ Dynamic Anisotropy in MHD Turbulence induced by mean magnetic field}


\author{Sita Sundar$^{1}$, Mahendra K. Verma$^{2}$,  Alexandros Alexakis$^{3}$, and Anando G. Chatterjee$^{2}$}

\affiliation{$^{1}$Christian-Albrechts-Universit{\"a}t zu Kiel - 24098, Germany \\
$^{2}$Department of Physics, Indian Institute of Technology Kanpur- 208016, India \\
$^{3}$Laboratoire de Physique Statistique, ENS Paris- 75230, France }

\begin{abstract}
 In this paper we study the development of anisotropy in strong MHD turbulence in the presence of 
a large scale magnetic field $B_0$ by analyzing the results of direct numerical simulations.  
Our results show that the developed anisotropy among the different components of the velocity 
and magnetic field is a direct outcome of the inverse cascade of energy of the perpendicular 
velocity components $\bf u_\perp$ and a forward cascade of the energy of the parallel component $\bf u_\parallel$.
The inverse cascade develops for strong ${B_0}$ where the flow exhibits strong vortical structure by the suppression
of fluctuations along the magnetic field. Both the inverse and the forward cascade are examined in detail by investigating
the anisotropic energy spectra, the energy fluxes and the shell to shell energy transfers among different scales.
 \end{abstract}
\maketitle

\section{Introduction}

Magnetohydrodynamics (MHD) provides the macroscopic equations for the motion of a conducting fluid that is coupled with the electrodynamics equations. 
MHD flows are ubiquitous in nature, and they are observed in the interstellar medium, galaxies, 
accretion disks, star and planet interiors, solar wind, Tokamak etc.
In such flows, the kinetic Reynolds number $Re$ (defined as $Re=UL/\nu$, 
where $U$ is the rms velocity, $L$ is the domain size, and $\nu$ is the kinematic viscosity)
and magnetic Reynolds number $Rm$ (defined as $Rm=UL/\eta$,  where $\eta$ is the 
magnetic diffusivity) are so large that the 
flows are turbulent with a large continuous range of excited scales, from the largest scales 
where energy is injected to the smallest scales where energy is dissipated.
Furthermore, in most of these systems, reasonably strong magnetic fields are known to exist,
with correlation lengths much larger than those of the turbulent flow.
These large-scale magnetic fields present in these systems induce dynamic anisotropy, and  hence play significant dynamical role in the flow evolution.

Resolving both the large scale magnetic fields and the small scale turbulence  by direct numerical simulations 
is still a major challenge even with the presently available supercomputers (see \cite{Alexakis2013}). 
One of the possible simplifications around this difficulty is to model the large-scale magnetic 
field by a uniform magnetic field ${B_0}$, and study its effect on the small scale turbulence. 
This approximation simplifies the analysis of the system as it allows to study the effect of large magnetic 
fields on small scale turbulence without tracking down their slow evolution.
For example, various features of the solar corona (e.g.,  the magnetic structures associated with prominence, coronal holes with their open field lines, and coronal loops) are modeled using such a ``magnetofluid with mean ${B_0}$ field'' approximation.
Other systems of interest where such an approximation is advantageous include the solar wind, where the inertial-range
fluctuations are subjected to a mean magnetic field, 
and fusion devices, like ITER, that involve large toroidal magnetic fields.

MHD turbulence in the presence of a mean magnetic field has been the subject of many 
studies ~\citep{Iroshnikov1963,Kraichnan1965,Shebalin1983,Zank1993,Oughton1994}.
The initial phenomenological estimates for the energy spectrum $E(k)$ based on Alfv\'{e}n effects and isotropy lead to the prediction
of an energy spectrum $E(k)\propto k^{-3/2}$ \cite{Iroshnikov1963,Kraichnan1965}.  
Verma~\cite{Verma:PP1999,Verma2004} showed that the ``random" large-scale mean magnetic field ${B_0}$ 
gets renormalized to yield $B_0(k) \sim k^{-1/3}$ and Kolmogorov-like energy spectrum ($E(k) \sim k^{-5/3}$).  
This result is also consistent with energy spectrum derived by re-normalizing viscosity and 
resistivity~\cite{Verma:PRE2001}.  

The presence of a large-scale mean magnetic field however supports propagation of Alfv\'{e}n waves that makes the flow  anisotropic.  
The first studies of anisotropy by Shebalin {\em et. al.}~\cite{Shebalin1983} in two-dimensional magnetohydrodynamics and by 
Oughton {\em et al.}~\cite{Oughton1994} in three dimensions quantified the anisotropy by measuring the angles
\begin{equation}
\theta_{u,b} = \tan^{-1} \frac{\sum_k k_z^2 E_{u,b}({\bf k})}{\sum_k (k_x^2+k_y^2) E_{u,b}({\bf k})}.
\end{equation}
where $E_{u,b}$ is the velocity or magnetic field energy spectrum, and $\hat{z}$ is the direction of the mean magnetic field.  
In their  low-resolution simulations ($k_\mathrm{max}=32$), they employed ${B_0} = 0$ to $16$, and
showed that strong anisotropy arises due to the mean magnetic field with the anisotropy being strongest at higher wavenumbers and thus 
it can not be neglected.
Phenomenological theories that take in to account anisotropy predict that the anisotropic energy spectrum scales  as $k_\perp^{-5/3}$~\cite{Goldreich1995}  (where $k_\perp$ is 
the wave number perpendicular to the mean magnetic field) or as $k_\perp^{-3/2}$~\cite{Boldyrev2006}.  
Simulations of Boldyrev {\em et al.}~\cite{Boldyrev:PhRv2009,Boldyrev2012,Boldyrev:ApJ2014} support $-3/2$ exponent, while those by 
Beresnyak~\cite{Beresnyak:ApJ2009,Beresnyak:PhRvL2011,Beresnyak:ApJ2014} argue in favour of Kolmogorov's exponent $-5/3$.  Thus, at present there
is no consensus on the energy spectrum for the MHD turbulence. 

The only case that analytical results have been derived is the weak turbulence limit where the uniform magnetic field is assumed to be very strong.   
In this limit, the evolution of the energy spectrum can be calculated analytically using an asymptotic expansion \citep{Galtier2000} that leads to 
the prediction $E(k_\perp)\propto k_\perp^{-2}$.   The predictions above however are valid only in  large enough domains in which many large-scale 
modes along the mean magnetic field exist.  In finite domains one finds an even  richer behavior.  
It has been shown~\citep{Alexakis2011,Reddy:POF2014,Reddy:POP2014} with the use of numerical simulations that in finite domains, three-dimensional 
MHD flows become quasi-two-dimensional for strong external magnetic field.    These states  exhibit high anisotropy with very weak variations along 
the direction of the magnetic field and resembles two-dimensional turbulence. In fact, it can be shown that for  ${B_0}$ above a critical value, 
the aforementioned two-dimensionalisation becomes exact \cite{Gallet2015}, with  three-dimensional perturbations dying off  exponentially in time. 
At intermediate values of ${B_0}$, however, three-dimensional perturbations are present and control the forward cascade of energy.

The degree of anisotropy in such quasi two-dimensionalized situations has been studied more recently.   To quantify scale-by-scale anisotropy, 
Alexakis {\em et al.}~\cite{Alexakis2007, Alexakis2011} partitioned the wavenumber space into coaxial cylindrical domains aligned along  the mean magnetic field direction, 
and into planar domains transverse to mean field.  Using this decomposition, Alexakis~\cite{Alexakis2011} studied the energy spectra and fluxes, as well as 
two-dimensionalization of the flow for mean magnetic field strengths ${B_0} = 2$, $5$, and $10$.   He reported an inverse energy cascade for the wavenumbers 
smaller than the forcing wavenumbers. Teaca {\em et al.}~\cite{Teaca:PRE2009} decomposed the spectral space into rings, and  arrived at similar conclusion as above.  
Teaca {\em et al.} observed that the energy tends to concentrate near the equator strongly as the strength of the magnetic field is increased.  
They also showed that the constant magnetic field facilitates  energy transfers from the velocity field to the magnetic field.  
In the present paper, we study in detail the development of anisotropy in such flows and relate it to the development of the 
inverse cascade.

The outline of the paper is as follows. We introduce the theoretical framework in Sec. II followed by details
of the numerical simulations in Sec. III. Next, we discuss  the anisotropic spectra  in Sec. IV, and energy transfers diagnostics like energy flux and shell-to-shell energy transfers in Sec. V. Finally, we conclude in section VI.

\section{Setup and governing equations}   
We consider an incompressible flow of a conducting fluid  in the presence of a constant and strong guiding magnetic field ${\bf B_0}$ along $\hat{z}$ direction.   The incompressible MHD equations \citep{Roberts1967,Verma2004} are given below:
\begin{eqnarray}
&\frac{\partial }{\partial t}{\bf u} + ({\bf u} \cdot \nabla)  \mathbf{u}  =-\nabla P+ ({\bf B} \cdot \nabla)  \mathbf{b} + \nu {\nabla}^2 \mathbf{u}  + \mathbf{f} \nonumber \\
&\frac{\partial }{\partial t}{\bf b} + ({\bf u} \cdot \nabla)  \mathbf{b}  = ({\bf B} \cdot \nabla)  \mathbf{u} + \eta \nabla^2 \mathbf{b} \\
&\nabla \cdot \mathbf{u}=0, \qquad  \nabla \cdot \mathbf{b}=0. \nonumber
\label{incom_mhd1} 
\end{eqnarray}
Here ${\bf u}$ is the velocity field, ${\bf B}$ is the  magnetic field, ${\bf f}$ is the external forcing,  $P$ is the total (thermal + magnetic) pressure, $\nu$ is the viscosity, and $\eta$ is the magnetic diffusivity of the fluid.  We take $\nu=\eta$, thus the magnetic Prandtl number $Pm = \nu/\eta$ is unity.   The total magnetic field is decomposed into its mean part ${ B_{0}}\hat{z}$ and the fluctuating part ${\bf b}$, i.e. ${\bf B} = { B_0} \hat{z}+ {\bf b}$.  Note that in the above equations, the magnetic field has the same units as the velocity field.   

The above equations were solved using a parallel pseudospectral parallel code { \sc Ghost} \cite{ghost}
with a grid resolution $512^3$ and a fourth order Runge-Kutta method for time stepping.   The simulation box is of the size $(2\pi)^3$ 
on which periodic boundary condition on all directions were employed.  The velocity field was forced randomly at the intermediate wavenumbers  
satisfying $8 \le |k| \le 10 $. This allowed to observe the development of both the inverse cascade and the forward cascade when they are present.  
The simulations were evolved for sufficiently long times so that either a steady state was reached, or until we observe dominant energy at the largest scales 
due to the inverse cascade of energy (for large ${B_0}$). In the simulations the forcing amplitude  was controlled, while the saturation level of the kinetic energy  
is a function of the other control parameters of the system.  Thus, the more relevant non-dimensional control parameter  is the Grasshof number defined as 
$G\equiv \|{\bf f}\| L^3 / \nu^2$, where $\|\cdot\|$ stands for the $L_2$ norm, and $L=2\pi$ is the length scale of the system.  Alternatively, we can use 
the Reynolds number $Re=\| {\bf u}\|L/\nu$ based on the rms value of the velocity. Note however  that $Re$ evolves in time in the presence of an inverse cascade.  
For further details of simulations, refer to Alexakis  \cite{Alexakis2011}.

We examine two different values of ${B_0=2}$ and $10$.
The results of these simulations were first presented in \cite{Alexakis2011} and correspond to the runs R2 and R3 respectively in that work.
The values of the control parameters used and of the  basic observable are summarized in table \ref{param_table}. 
The runs have relatively moderate Reynolds number due to the forcing at intermediate wavenumbers.  Therefore we do not focus on the energy spectra.  
Rather we aim  to unravel the mechanisms that lead to the redistribution of energy and development of anisotropic turbulence due to the mean magnetic field.

\begin{table}[h]
\caption{Steady-state  parameters of the simulation:  Grasshof number $G\equiv \|{\bf f}\| L^3 / \nu^2$, Reynolds number $\|{\bf u}\| L/\nu$, kinetic and magnetic energies,  ${B_0}/\|{\bf u}\|$, $r_A^{-1} = \|{\bf b}\|^2/\|{\bf u}\|^2$, kinetic and magnetic dissipation rates, anisotropic parameters $A_u$ and $A_b$(see Eq.~(\ref{aniso1})). 
The values are obtained from single snapshots  and not by time-averaging.}
\label{param_table}
\setlength{\tabcolsep}{7 pt}
\hspace*{-1cm}
\begin{tabular}{ c c c c c  c  c  c  c c c  }
\toprule[1.2pt]
$Gr^{1/2}$ & $Re$  & ${B_0}$ &  $\|{\bf u}\|^2$ &  $\|{\bf b}\|^2$ &${B_0}/\|{\bf u}\|$  & $r_A^{-1}$  & $\nu\|{\bf \nabla u}\|^2$ & $\eta\|{\bf\nabla b}\|^2$  &   $A_u$  & $A_b$ \\
\hline
2500       & $1.09 \times 10^4$ & 2     &     0.24      &       0.18   &4.08 & 0.75 &   0.043          &   0.041             &  0.53  & 0.73 \\
2500       & $1.53\times 10^4$ & 10    &     0.47       &       0.012  & 14.6 & 0.026 & 0.015          &   0.0021            &  3.7  & 1.6 \\
\bottomrule[1.2pt]
\end{tabular}
\end{table}

In later sections, we  analyze  the anisotropic energy spectra and energy transfer diagnostics using the generated numerical data 
by employing another pseudo-spectral code {\sc Tarang}~\cite{tarang}. We describe the anisotropic energy spectra, as well as the fluxes and the energy transfers 
involving the velocity and magnetic fields, generated during the evolved state.  Throughout the paper, we denote $u_{\parallel} = u_{z}$ and ${\bf u}_{\perp}= (u_{x}, u_{y})$.

\section{Spectra and anisotropy} 

\begin{figure}[h!]
\hspace*{-1.5cm}
\includegraphics[scale=.7,  trim =  0cm 15.5cm 0cm 4cm, clip =true]{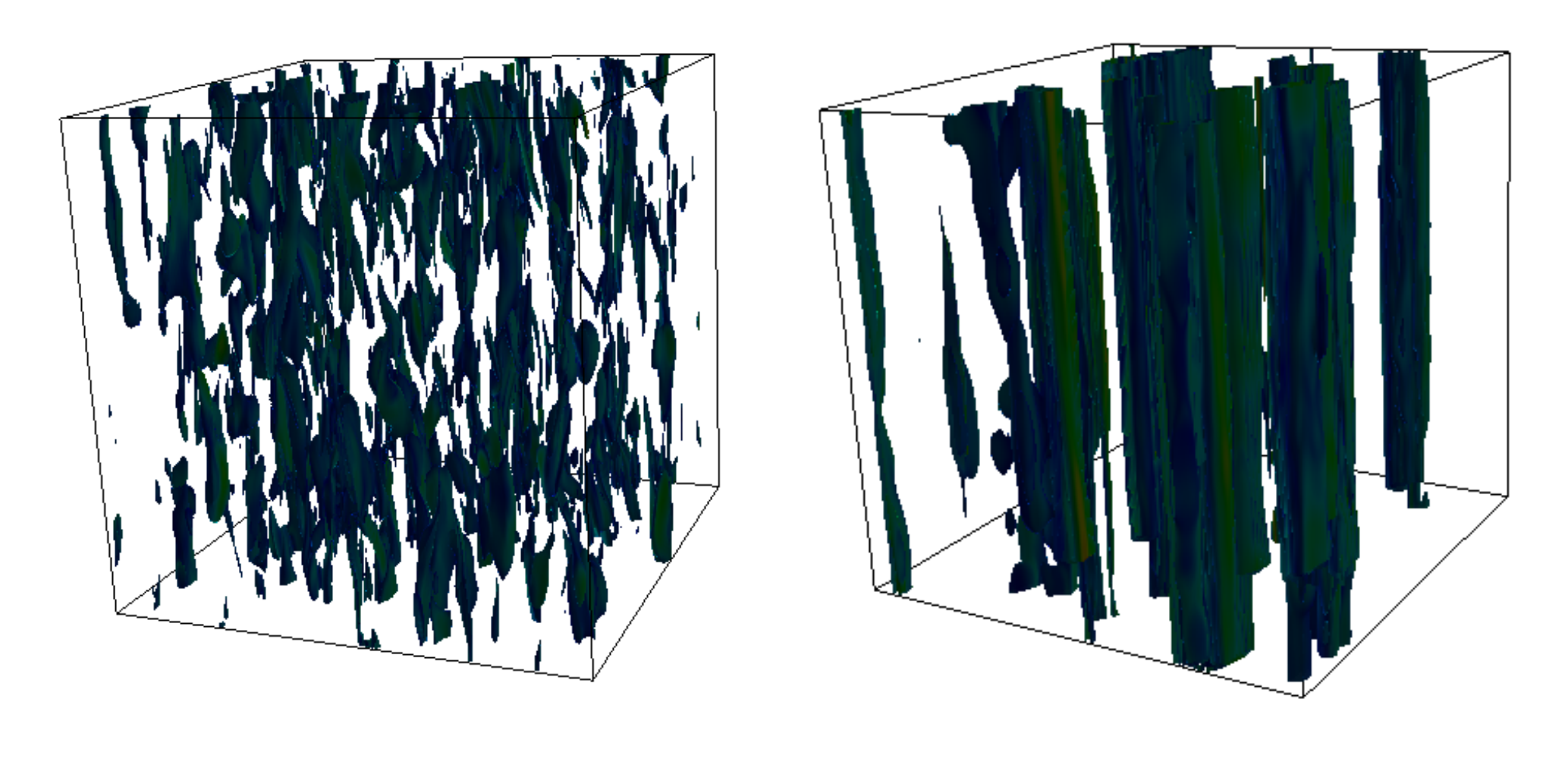}\\
%
\vspace{-0.8cm}
\hspace*{-1.2cm}(a)  \hspace*{9.2cm} (b)
\caption{Isosurfaces of magnitudes of vorticity $|\boldsymbol{\omega}|$ for  mean magnetic field
(a) ${B_0=2}$  and (b) ${B_0=10}$.}
\label{fig:vorticity} 
\end{figure}

\begin{figure}[h!]
\hspace*{-1.0cm}
\includegraphics[scale=0.25,  trim =  0cm 3cm 0cm 5cm, clip =true]{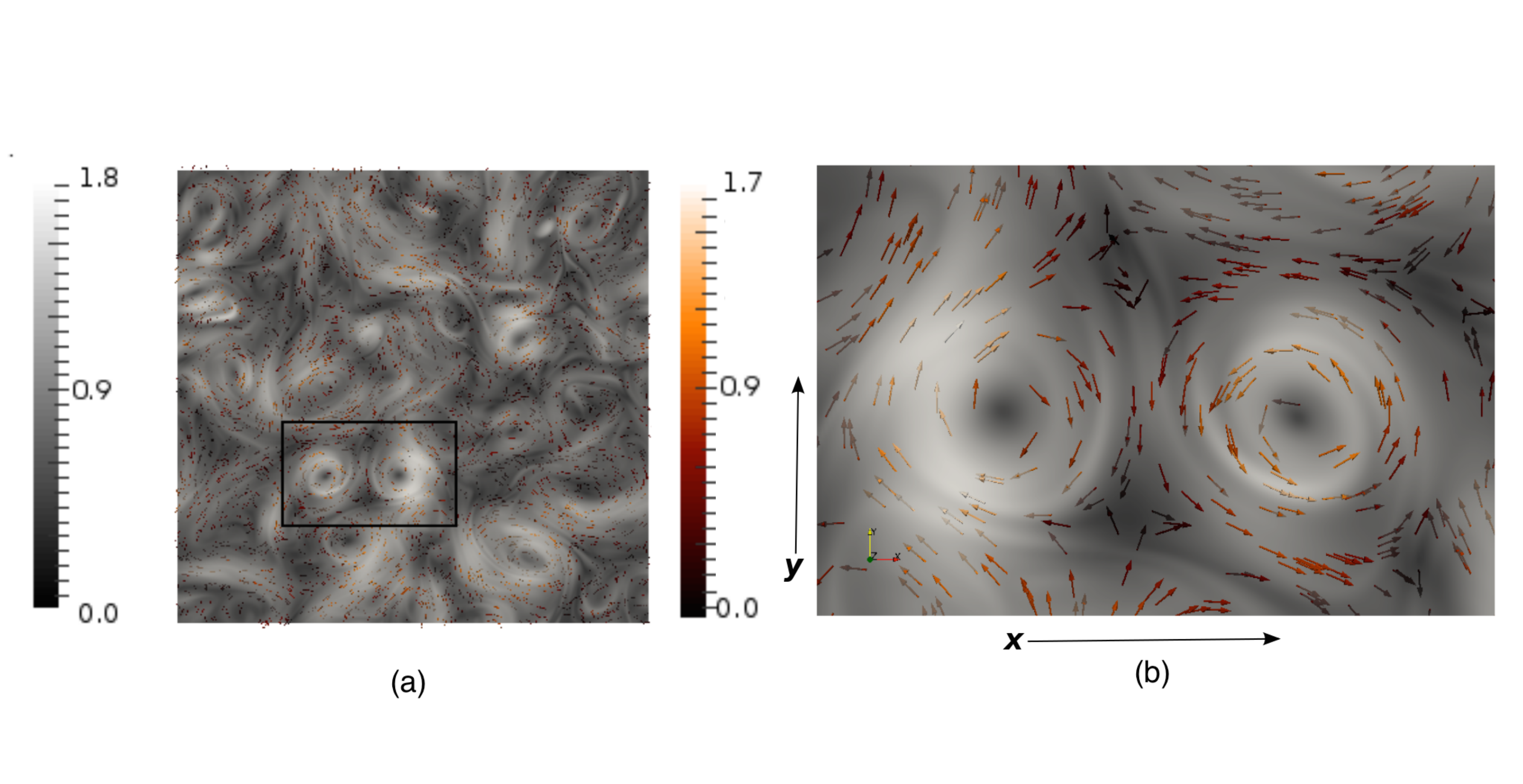}
\caption{For ${B_0=10}$, a horizontal cross-sectional view of (a) density plot of  $|\boldsymbol{\omega}|$ (arrows) along with the velocity vectors (gray background), The `grayscale' and `hot-cold' (shown by `dark red/brown') colorcode correspond to the magnitude of velocity field and vorticity respectively. (b)  A zoomed view of area inside black rectangle of subplot (a). }
\label{fig:vort_cont} 
\end{figure}

First we present visualizations of the two examined flows for ${B_0=2}$ and $10$ to demonstrate the anisotropy of the flow.  In Figure~\ref{fig:vorticity},  
we present the iso-surfaces of the magnitude of the vorticity $|\boldsymbol{\omega}|$, where $\boldsymbol{\omega} = \nabla \times {\bf u}$.  The flow has  vortical columnar structures along ${B_0}$ that becomes stronger as ${B_0}$ is increased.  To get further details of the flow structure, we make a horizontal section for the ${B_0=10}$ case. 
In Figure~\ref{fig:vort_cont}(a) we show the density plot of vorticity magnitude   along with velocity vectors $(u_{x}, u_{y})$.
The flow develops strong vortical structure, with strong 
 $u_y$ and $u_x$ components, while modes that vary along $\hat{z}$ are very weak. The reason for the
 formation of these structures is discussed in detail in Sec.~IV).

To quantify the anisotropy of the flow, we  propose  anisotropy measures $A_{u}$ and $A_{b}$ for the velocity and magnetic fields  as
\begin{equation}
A_{ u} = \frac{E_{u}^\perp}{2E_{u}^\parallel};~~~ A_{b} = \frac{E_{b}^\perp}{2E_{b}^\parallel}.
\label{aniso1}
\end{equation}
where $E_u^\perp=\left\langle u_x^2+u_y^2 \right\rangle/2$ and  $E_u^\parallel=\langle u_z^2\rangle /2$, where the angular brackets stand for spatial average.    The quantities $E_u^\perp$ and $E_u^\parallel$ represent the kinetic energies of the perpendicular and parallel components of the velocity field. Similar definitions are employed for the magnetic field.   
The anisotropy parameter $A_{u,b}$ measures the degree of anisotropy among the different components of the velocity and magnetic field.
It is defined such that $A_{u,b} = 1$ for isotropic flow with  
$\left\langle u_x^2\right\rangle = \left\langle u_y^2 \right\rangle= \left\langle u_z^2\right\rangle $, but it deviates from unity for anisotropic flows.  
In Table \ref{param_table}, we list $A_{u}$ and $A_{b}$ for the two runs.
For ${B_0}=2$,  both $A_u$ and $A_b$ are smaller than unity, i.e. $E^\perp_u < 2E^\parallel_u$
(due to the particular choice of forcing used),
while for ${B_0}=10$, their magnitude is substantially higher than unity  ($E_\perp > 2E_\parallel$)
that as we shall show later is due to the presence of an inverse cascade:  
the flow is quasi two-dimensional, hence it  exhibits strong inverse cascade of kinetic energy leading to buildup of kinetic energy at 
large scales.

\begin{figure}[h!]
\includegraphics[scale=0.9,  trim = 0cm 6cm 0cm 0cm, clip =true]{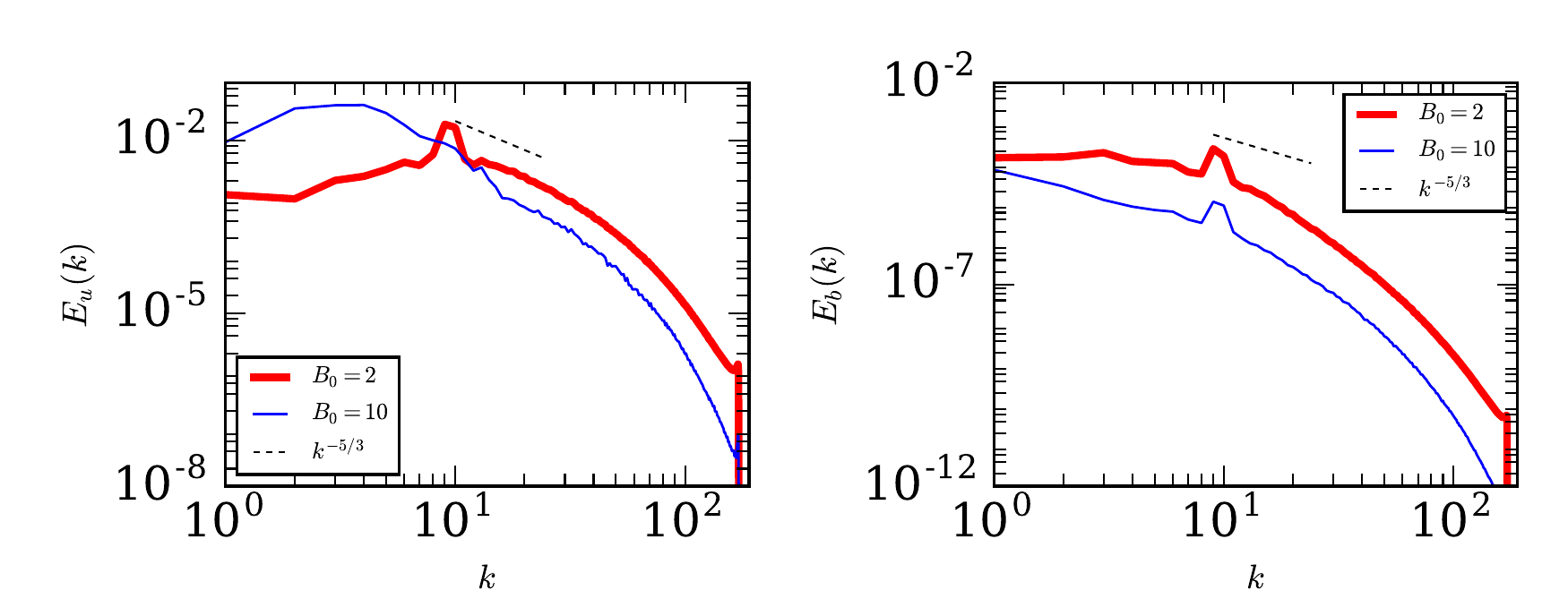}
\hspace*{2.0cm}(a)  \hspace*{7.5cm} (b)
\caption{Plots of (a) Kinetic energy spectrum, $E_u(k)$ and (b) Magnetic Energy Spectrum, $E_b(k)$   for  ${B_0}=2$ and $10$.}
\label{fig:enspectra}
\end{figure}

Further insight can be obtained by studying  the distribution of energy among the different components and different modes in the Fourier space.
For isotropic flows, the energies of all the modes and all components within a thin spherical shell in Fourier-space are statistically equal. 
Hence, sum  of the energies of all the Fourier modes in a spherical shell of radius $k$ is often reported as one-dimensional energy spectrum $E(k)$.  It provides information about the distribution of energy at different scales.  The one-dimensional spectra for the velocity and the magnetic field are  shown in Fig.~\ref{fig:enspectra}.   For the ${B_0} = 10$ case,  the kinetic energy peaks at the large scales while the magnetic fluctuations are suppressed. 
This is due to the presence of an inverse cascade of energy as discussed in \cite{Alexakis2011} (further discussions in Sec.~V).   For ${B_0}=2 $ the inverse cascade is reasonably weak, if at all. 
This is also consistent with the values of $A_u$ and $A_b$ (presented in Table I) for the two cases and is discussed in detail in Secs.~IV-V.  The dashed line indicates the $k^{-5/3}$ power-law scaling; our inertial range is too short to fit with this spectrum.  As discussed in the introduction in this paper, our attempt is not to differentiate  between the exponents $-3/2$ and $-5/3$,  but rather study the effects of large $B_0$ on the global statistics of the flow.  

\begin{figure}[h!]
\hspace*{-1.0cm}
\includegraphics[scale=1,  trim = 0cm 6cm 0cm 0cm, clip =true]{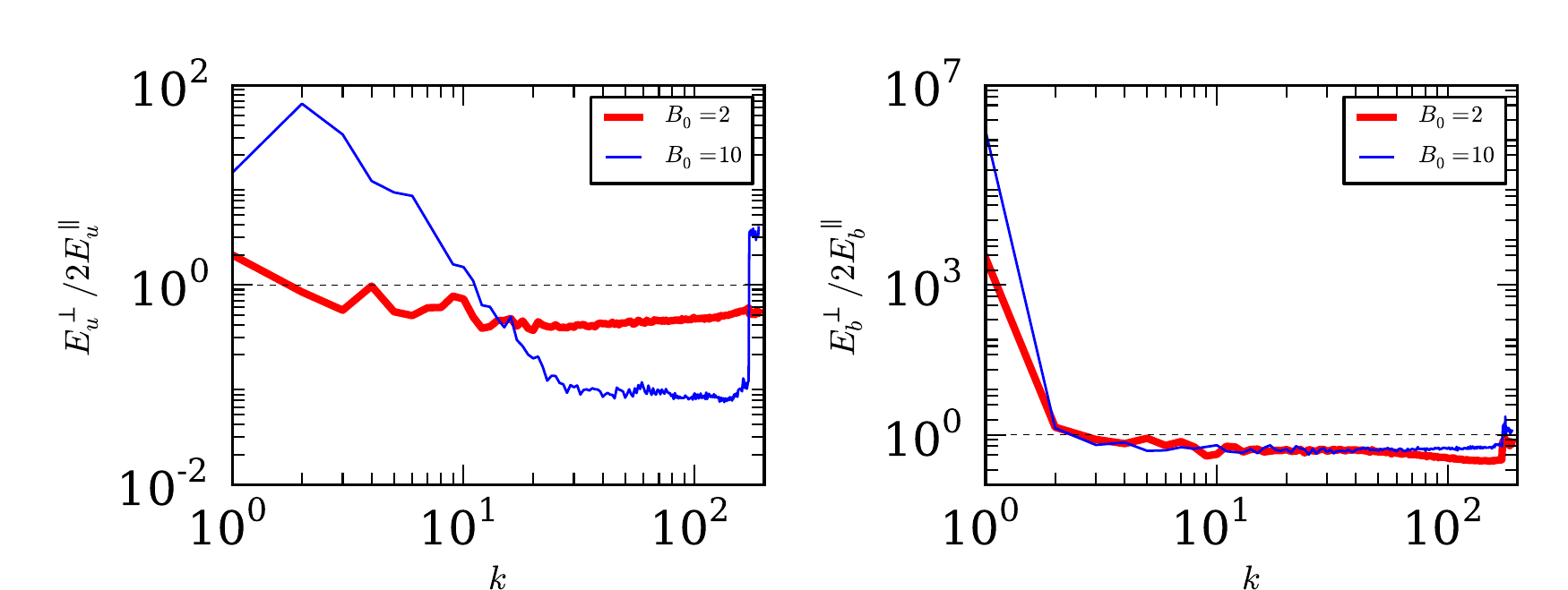}
\hspace*{2.0cm}(a)  \hspace*{7.5cm} (b)
\caption{Plots of anisotropy spectrum of the velocity field $A_{u}(k) = \frac{E_{u}^\perp (k)}{2E_{u}^\parallel(k)}$ and magnetic field $A_{b}(k) = \frac{E_{b}^\perp(k)}{2E_{b}^\parallel(k)}$.}
\label{fig:anisotropy}
\end{figure}
To explore the nature of the anisotropy at different length scales, we work in Fourier space, in which the equations are
\begin{eqnarray}
\left(\frac{d}{dt}  + \nu k^2 \right) {u}_i ({\bf k})-i({\bf B}_0 \cdot {\bf k}){b}_i ({\bf k}) &  = &  - i k_i P({\bf k}) -i k_j \sum_{{\bf k=p+q}}{u}_j ({\bf q}) {u}_i ({\bf p} )    \nonumber  \\
& & + i k_j \sum_{{\bf k=p+q}}{b}_j ({\bf q}) {b}_i ({\bf p} )  +{\bf f}({\bf k}), \label{eq:uk}\\
\left(\frac{d}{dt}  + \eta k^2 \right) {b}_i ({\bf k}) -i({\bf B}_0 \cdot {\bf k}){u}_i ({\bf k}) &  = &    -i k_j \sum_{{\bf k=p+q}}{u}_j ({\bf q}) {b}_i ({\bf p} )  + i k_j \sum_{{\bf k=p+q}}{b}_j ({\bf q}) {u}_i ({\bf p} ), \label{eq:bk}
\end{eqnarray}
where $\hat{\bf u}({\bf k}), \hat{\bf b}({\bf k})$ are the Fourier transform of ${\bf u,b}$ respectively. First we compute  wavenumber-dependent anisotropy parameters:
\begin{equation}
A_{u}(k) = \frac{E_{u}^\perp (k)}{2E_{u}^\parallel(k)};~~~ A_{b}(k) = \frac{E_{b}^\perp(k)}{2E_{b}^\parallel(k)},
\label{aniso_k}
\end{equation}
where $E_{u}^\perp (k)$ represents sum of energy of  the Fourier transform of ${\bf u}_\perp$ in the shell $(k-1:k]$.  Similar definitions holds for other spectra. Fig.~\ref{fig:anisotropy}(a,b) exhibits the plots of $A_{u}(k)$ and $A_{b}(k)$ respectively.  For ${B_0}=2$ , $A_{u}(k) >1$ for $k=1$,  and $A_{u}(k) \approx 1/2$   for $k>1$.   However for ${B_0}=10$, $A_{u}(k)$ is strongly anisotropic with $A_{u}(k) \gg 1$ for $k<k_{f}$, but $A_{u}(k) \ll 1$ for $k > k_{f}$.   Thus, for ${B_0}=10$, the  two-dimensional components in the large-scale velocity field dominate,   consistent with the flow profile of Figs.~\ref{fig:vorticity} and \ref{fig:vort_cont}. Note that  $u_{\parallel}$ dominates over $u_{\perp}$ at large wavenumbers.   This behavior is very similar to anisotropic behavior in quasi-static MHD reported by Reddy and Verma~\cite{Reddy:POF2014} and Favier {\em et al.}~\cite{Favier2010}.

For magnetic field $\bf b$, $A_{b}(k)$ is very large for $k=1$, but $A_{b}(k) \sim 1$ for $1<k<k_f$, while it is less than unity for $k>k_f$.
The large peak at $k=1$ for the ratio $E_b^\perp/E_b^\|$ is caused not due to excess
of  $E_b^\perp$ energy but rather due to the almost absence of $E_b^\|$ in the large scales.
Indeed the quasi-2D motions of the flow are not able to amplify $E_b^\|$ and thus the ratio $A_b$
almost diverges at $k=1$. 
 For Alfvenic turbulence where there is only a forward cascade it is observed
that $|\delta B_\perp|^2  \gg  |\delta B_\parallel|^2$ (see ~\citep{0004-637X-753-2-107,0004-637X-674-2-1153}).
However in our case as we explain later in our text part of $E_u^\perp$ and $E_b^\perp$
cascades inversely while $E_u^\|$ and $E_b^\|$ cascade forward causing an excess of  $E_b^\|$
and  $E_u^\|$ in the small scales.
 
\begin{figure}[h!]
\includegraphics[scale=0.5,  trim = 3cm 8cm 0cm 8cm, clip =true]{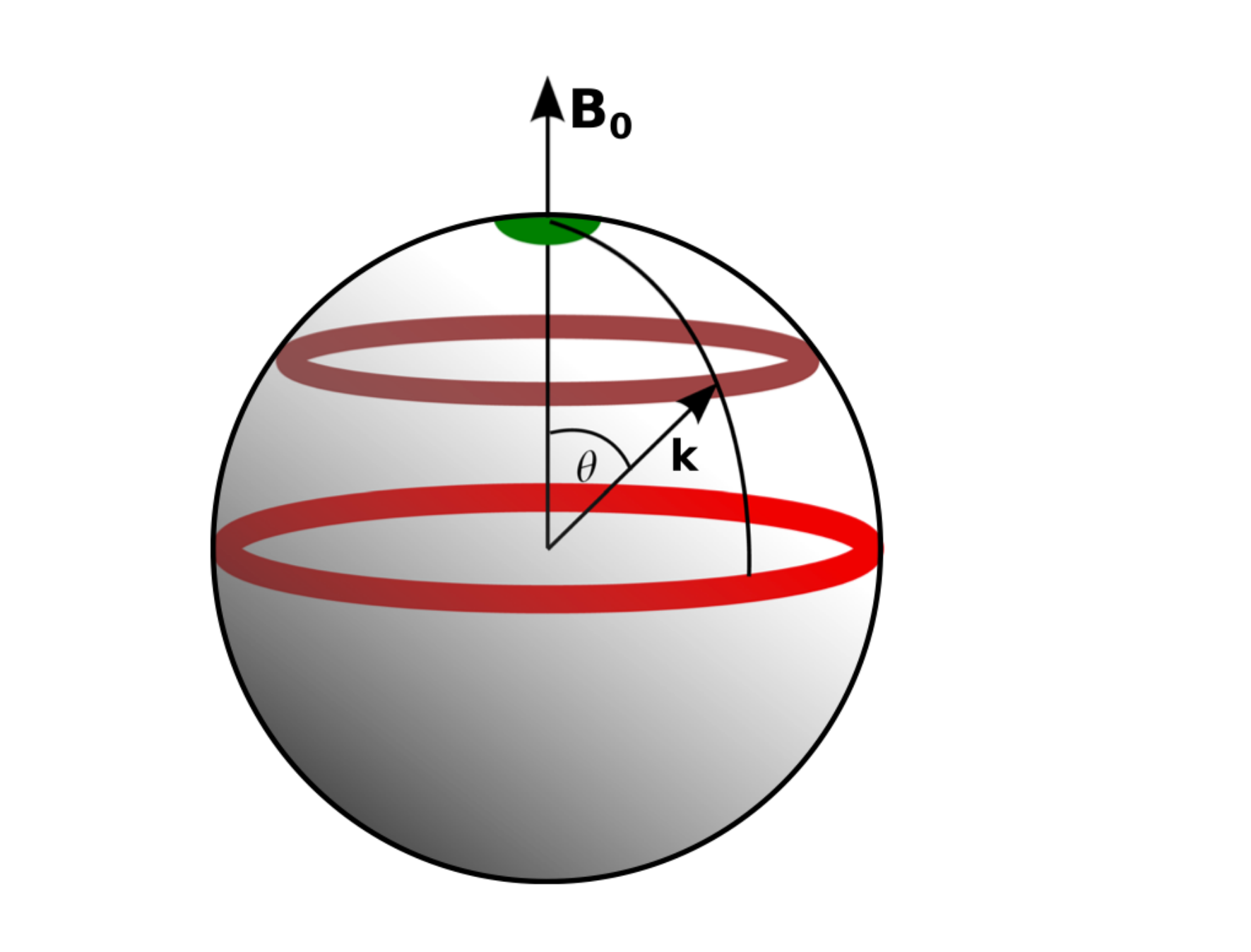}
\caption{Illustration of the ring decomposition in the spectral space. This figure is taken from Ref.~\cite{Reddy:POP2014}. [Reprinted with permission from Sandeep Reddy [K. S. Reddy, R. Kumar, and M. K. Verma, Physics of Plasmas 21, 102310 (2014). Copyright 2014, AIP Publishing].}
\label{fig:ring_schem}
\end{figure}

A different measure of anisotropy is provided by looking at the distribution of energy in spectral space
using a ring decomposition shown in Fig.~\ref{fig:ring_schem} that we now discuss.  
A spherical shell in Fourier space is divided into rings such that each ring is characterized by two indices---the shell index $k$, 
and the sector index $\beta$ \cite{Teaca:PRE2009,Reddy:POF2014}.  
The energy spectrum of a ring, called the {\em ring spectrum}, is defined as
\begin{equation}
E\left(k,\beta \right)=\frac{1}{C_{\beta}}\sum_{\substack{k-1 < k' \le k;\\
\zeta_{\beta-1} < \angle\left(\mathbf{k}'\right)  \le  \zeta_{\beta}}
}\frac{1}{2}\left|\mathbf{\hat{u}}\left(\mathbf{k}'\right)\right|^{2},
\label{eq:e_kbeta}
\end{equation}
where $\angle\mathbf{k}'$ is the angle between $\mathbf{k}'$ and the unit vector $\hat{z}$, and the sector $\beta$ contains the modes  between the angles  
$\zeta_{\beta-1}$ to $\zeta_{\beta}$. When $\Delta \zeta$ is uniform, the sectors near the equator contain more modes than those near the poles. 
Hence, to compensate for the above, we divide the sum $\sum_{k} |{\bf \hat{u}(k')}|^2/2 $ by the factor $C(\beta)$ given by
\begin{equation}
C_{\beta}=\left|\cos\left(\zeta_{\beta-1}\right)-\cos\left(\zeta_{\beta}\right)\right|.
\end{equation}

\begin{figure}[h!]
\hspace*{-1.0cm}
\vspace*{-2.0cm}
\includegraphics[scale=0.9,  trim = 1cm 5cm 0cm 5cm, clip =true]{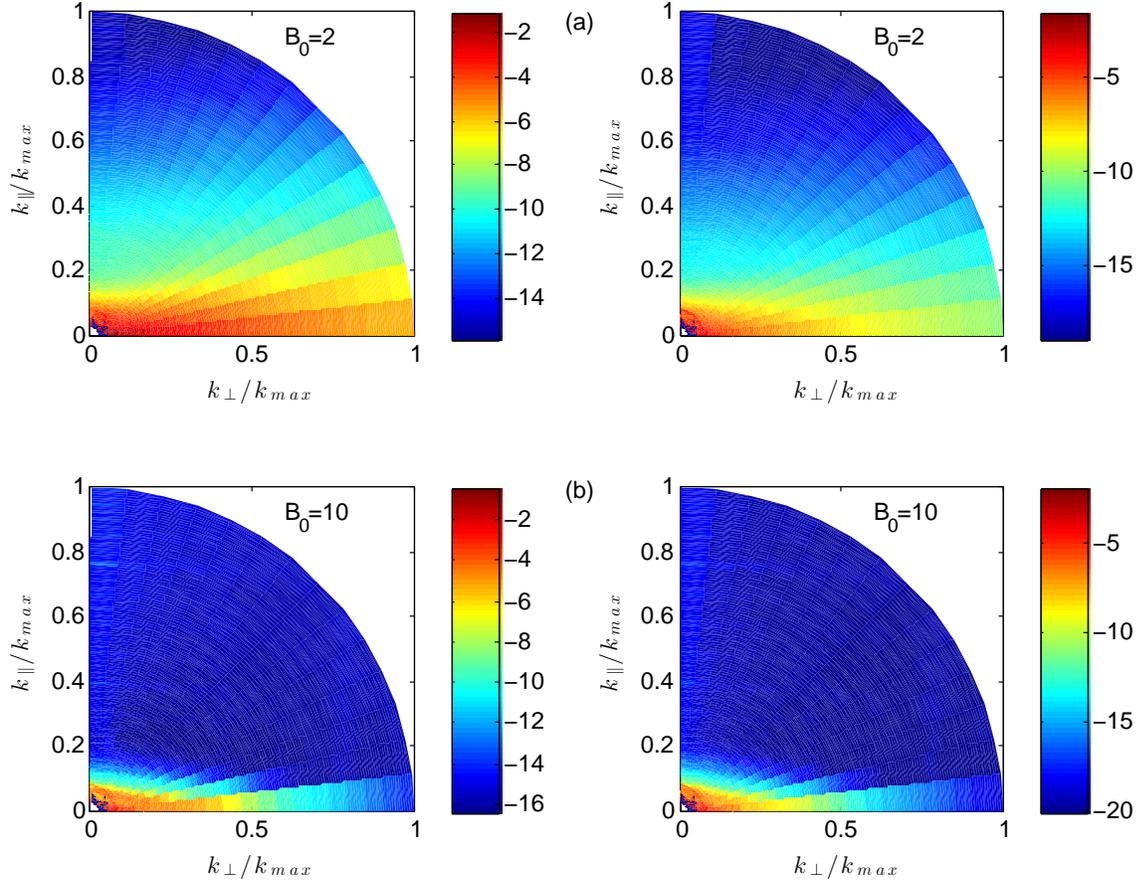}
\caption{The ring spectra in log scale:  $ \log(E_u(k,\theta))$ (left) and  $ \log(E_b(k,\theta))$ (right) for   (a)${B_0}=2$  and (b)${B_0}=10$.}
\label{fig:ringu}
\end{figure}

For the ring spectrum computations, we divide the spectral space in the ``northern" hemisphere into thin shells  of unit widths (see Eq.~(\ref{eq:e_kbeta})), which are further subdivided into 15 thin rings 
from $\theta = 0$ to $\theta = \pi/2$.  For the ring spectrum, we vary $k$ from 1 to $512\times (2/3) = 341$; the factor 2/3 arising due to aliasing. Taking  benefit of the $\theta \rightarrow (\pi - \theta)$ symmetry,  we do not compute the  energy of the rings in the ``southern" hemisphere.  
In Fig.~\ref{fig:ringu}, we show the density plots of the kinetic and magnetic ring spectrum $E(k,\beta)$ for ${B_0} = 2$ and $10$. 
From the plots it is evident that the kinetic and magnetic energy is stronger near the equator than the polar region, and the anisotropy increases with ${B_0}$.  
The anisotropy is greater for ${B_0}=10$, but the energy is concentrated near the equator even for ${B_0}=2$.

\begin{figure}[h]
\hspace*{-1.cm}
\vspace*{-0.5cm}
\includegraphics[scale=1,  trim = 0cm 0cm 0cm 0cm, clip =true]{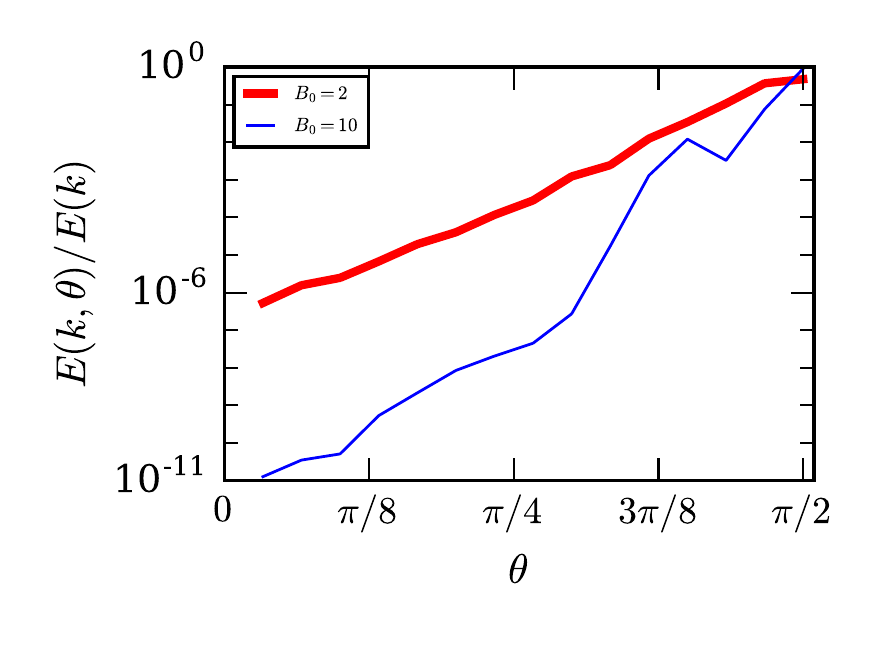}  
\caption{Plot of $E(k=20,\theta)/E(k=20)$ vs. $\theta$ for (a)${B_0}=2$ (thick line) and (b)${B_0}=10$ (thin line).}
\label{fig:ringN}
\end{figure}

For further illustration, in Fig.~\ref{fig:ringN} we show the normalized ring spectra  $E(k,\theta)/E(k=20)$ vs. $\theta$ for ${B_0} = 2$ and $10$ for  $k = 20$, which is a generic wavenumber in the inertial range.  Clearly $E(k,\theta)$, which is strongest for $\theta = \pi/2$, deviates strongly from a constant value, indicating anisotropy of the flow.  The deviation is stronger for ${B_0}=10$ than $B_0=2$, which is consistent with the earlier discussion.

\section{Energy flux and shell-to-shell energy transfers}

In this section we will study energy transfers that provide insights into the two-dimensionalization process in MHD turbulence.    To delve into the anisotropy of the flow and its causes, we investigate  the energy flux and energy exchange between the perpendicular and parallel components of the velocity field.  Earlier, energy transfers in the Fourier space have been studied in detail  by various groups  \cite{Dar2001,Verma2004,Alexakis2005,Debliquy2005}.   
Herein, we present an in-depth investigation of the energy transfers with comparatively stronger mean magnetic field amplitudes.

In hydrodynamics, for a basic triad of interacting wave-numbers $({\bf k},{\bf p},{\bf q})$ that satisfy ${\bf k}={\bf p}+{\bf q}$, the mode-to-mode energy transfer rate from the mode {\bf p} to the mode {\bf k} via mediation of the mode {\bf q}  is given by
\begin{equation}
S({\bf k} \mid {\bf p} \mid {\bf q}) = \Im\{[{\bf k} \cdot {\bf \hat{u}({\bf q})} ][{\bf \hat{u}}({\bf p}) \cdot {\bf  \hat{u}}^{*} ({\bf k})]\},
\end{equation}
where $\Im$ and $*$ denote respectively the imaginary part and complex conjugate of a complex number.
To investigate the energy transfer rate from a set of wave numbers $\mathcal{D}_p$ to a set of wave numbers $\mathcal{D}_k$ we  sum over  all the possible triads ${\bf k}={\bf p}+{\bf q}$:
\begin{equation}
\label{transf}
 \mathcal{T}(\mathcal{D}_k,\mathcal{D}_p) =  \sum_{{\bf k} \in \mathcal{D}_k} \sum_{{\bf p} \in \mathcal{D}_p} S({\bf k} \mid {\bf p} \mid {\bf q})
                                          = -\int {[\bf u_k (u\cdot\nabla)] u_p} dx^3  
\end{equation}
where $\bf u_k(x),u_p(x)$ express the velocity field filtered so that only the modes in  $\mathcal{D}_k,\mathcal{D}_p$ are kept respectably.  The energy flux $\Pi(k_0)$ then can be defined as  the  rate of energy transfer from the set $\mathcal{D}_s$ of modes inside a sphere of radius $k_0$ to modes outside the same sphere, i.e., 
\begin{equation}
 \Pi(k_0) =   \sum_{k<k_{0}} \sum_{p \ge k_{0}} S({\bf k} \mid{\bf p} \mid {\bf q}).
\end{equation}
Similarly we can define the
shell-to-shell energy transfer rate $T_n^m=\mathcal{T}(\mathcal{D}_n,\mathcal{D}_m)$  as the energy transfer rate from the modes in a spherical shell $\mathcal{D}_m$  to the modes in the spherical shell $\mathcal{D}_n$.

 MHD turbulence has six kinds of energy fluxes, namely the energy flux from inner u-sphere to outer u-sphere ($\Pi^{u<}_{u>}(k_0)$),  energy flux from inner u-sphere to outer b-sphere ($\Pi^{u<}_{b>}(k_0)$),  energy flux from inner b-sphere to outer b-sphere ($\Pi^{b<}_{b>}(k_0)$), energy flux from inner b-sphere to outer u-sphere  ($\Pi^{b<}_{u>}(k_0)$), energy flux from inner u-sphere to inner b-sphere ($\Pi^{u<}_{b<}(k_0)$), and energy flux from outer u-sphere to outer b-sphere  ($\Pi^{u>}_{b>}(k_0)$).  These fluxes can be computed using the following formulae~\cite{Verma2004,Dar2001,Alexakis2005,Debliquy2005,MininniPRE2005}:
\begin{eqnarray}
    \Pi^{u<}_{u>}(k_0) & = \, \sum_{k<k_{0}} \sum_{p \le k_{0}}  \Im\{[{\bf k} \cdot {\bf \hat{u}({\bf q})} ][{\bf \hat{u}}({\bf p}) \cdot {\bf  \hat{u}}^{*} ({\bf k})]\} = &  +\int {\bf u}^<_k {\bf (u\cdot\nabla)} {\bf u}^>_k dx^3, \nonumber\\
   \Pi^{u<}_{b>}(k_0) & =- \sum_{k<k_{0}} \sum_{p \le k_{0}}  \Im\{[{\bf k} \cdot {\bf \hat{b}({\bf q})} ][{\bf \hat{u}}({\bf p}) \cdot {\bf  \hat{b}}^{*} ({\bf k})]\} = &  -\int {\bf u}^<_k {\bf (b\cdot\nabla)} {\bf b}^>_k dx^3, \nonumber\\
 \Pi^{b<}_{u>}(k_0) & = -\sum_{k<k_{0}} \sum_{p \le k_{0}}  \Im\{[{\bf k} \cdot {\bf \hat{b}({\bf q})} ][{\bf \hat{b}}({\bf p}) \cdot {\bf  \hat{u}}^{*} ({\bf k})]\} = &  -\int {\bf b}^<_k {\bf (b\cdot\nabla)} {\bf u}^>_k dx^3, \nonumber\\ 
     \Pi^{b<}_{b>}(k_0) &= \, \sum_{k<k_{0}} \sum_{p \le k_{0}}  \Im\{[{\bf k} \cdot {\bf \hat{u}({\bf q})} ][{\bf \hat{b}}({\bf p}) \cdot {\bf  \hat{b}}^{*} ({\bf k})]\}  =& +\int {\bf b}^<_k {\bf (u\cdot\nabla)} {\bf b}^>_k dx^3
    \label{partialflux}
\end{eqnarray}
where ${\bf u}^<_k,{\bf b}^<_k$ express the velocity and magnetic fields where only the modes inside a sphere of radius $k$ are kept while  ${\bf u}^>_k,{\bf b}^>_k$ express the velocity and magnetic fields where only the modes outside the same sphere are kept.  The total energy flux, which is the total energy transfer from the modes inside the sphere to the modes outside the sphere, is
\begin{equation}
\Pi(k_0) =   \Pi^{u<}_{u>}(k_0) + \Pi^{u<}_{b>}(k_0) + \Pi^{b<}_{u>}(k_0) + \Pi^{b<}_{b>}(k_0) .
\end{equation}

\begin{figure}[h]
\includegraphics[scale=0.8]{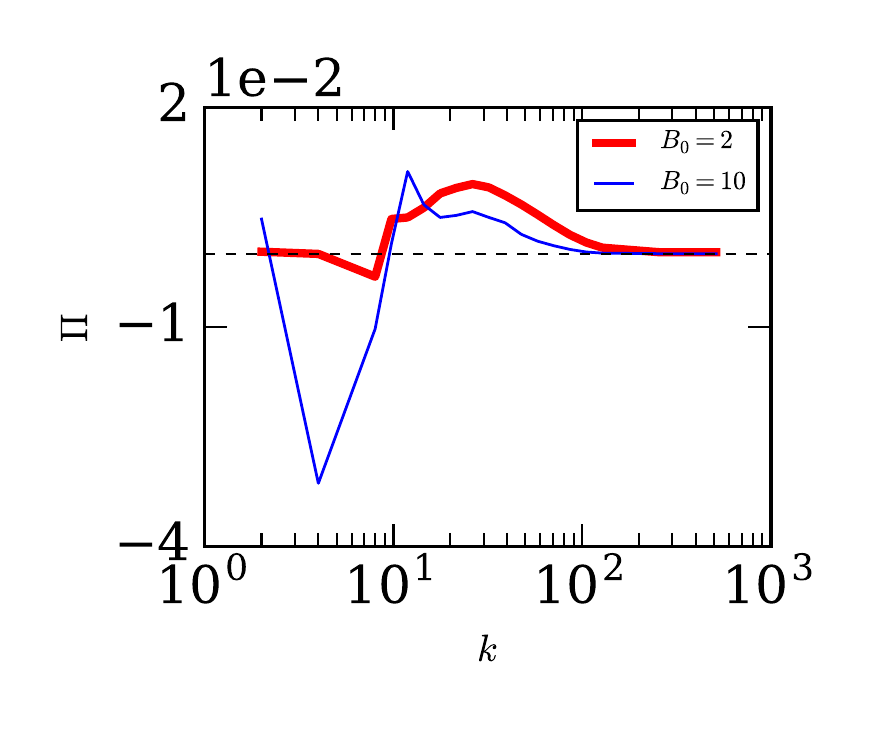}
\caption{Plot of total energy flux $\Pi(k)$ vs.~$k$.}
\label{fig:total_flux}
\end{figure}

 In the present paper, we compute the energy fluxes for 19 concentric spheres with their centres at k = (0, 0, 0).   The radii of the first three spheres are 2, 4, and 8, and those of the  last two spheres are  $512 \times 2/3 = 170.5)$ and $r_\mathrm{max} = 512 \times 2/3 = 341$.     Here the factor 2/3 is introduced due to dealiasing.  The intermediate shells are based on the powerlaw expression
\begin{equation}
r_i  =  r_3 \left[\frac{r_\mathrm{max}}{16.0}\right]^{\frac{i-3}{n-4}}.
\end{equation}
where $r_3=8$ is radius of the third sphere, $r_\mathrm{max}$ is the radius of the last sphere, and $n=19$ is the total number of spheres.  Hence, the radii of the spheres are   2.0, 4.0, 8.0, 9.8, 12.0, 14.8, 18.1, 22.2, 27.2, 33.4, 40.9, 50.2, 61.5, 75.4, 92.5, 113.4, 139.0, 170.5, and 341.0.
In the inertial range we bin the radii of the shells  logarithmically  keeping in mind the powerlaw physics observed here.   The inertial range however is too short since the forcing band is shifted to $k = [8,10]$. 

\begin{figure}[h]
\hspace*{-2.2cm}
\includegraphics[scale=0.7]{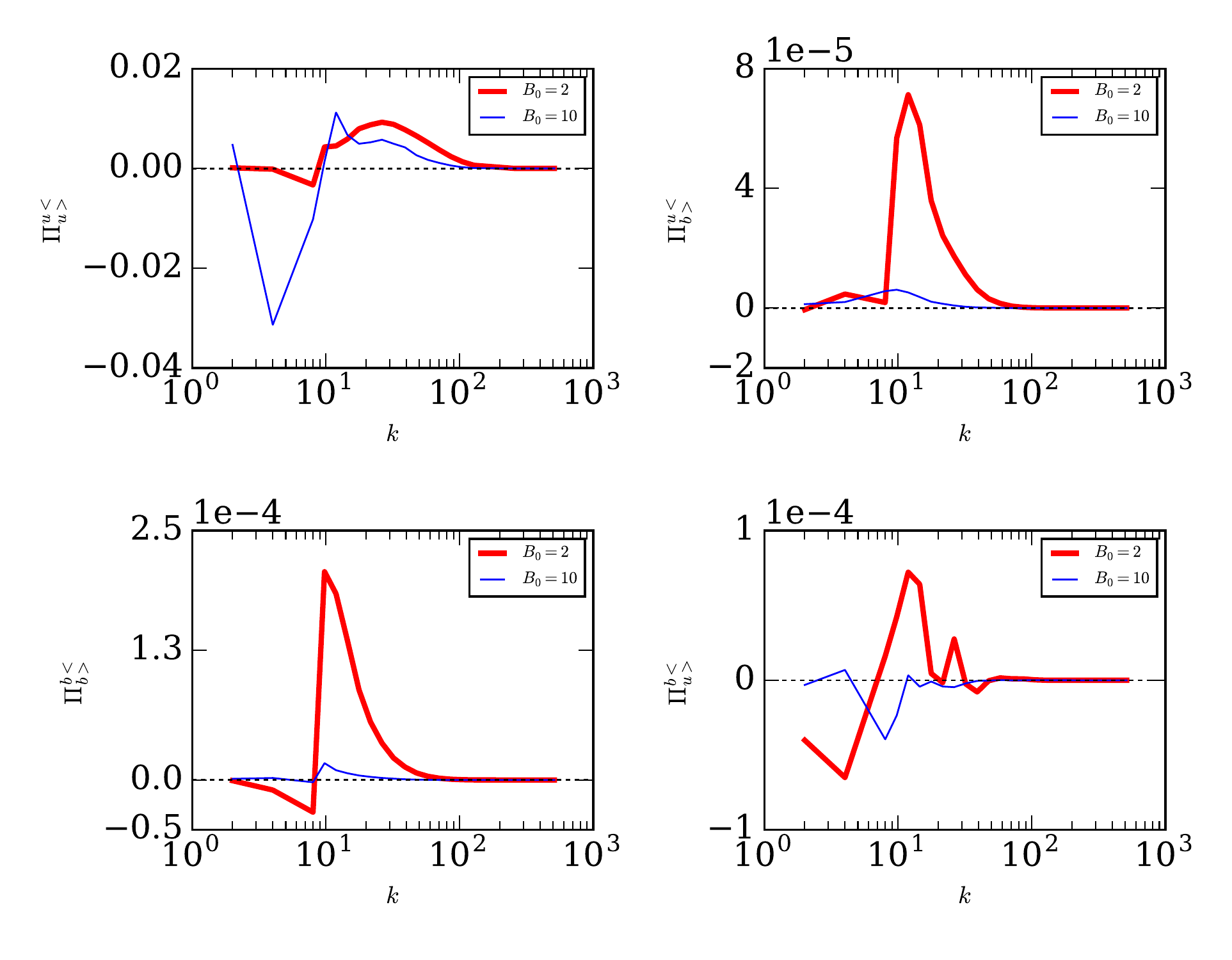}
\caption{Plots of energy fluxes $\Pi^{u<}_{u>}$,  $\Pi^{u<}_{b>}$,
$\Pi^{b<}_{b>}$, and  $\Pi^{b<}_{u>}$ {\it vs.  k}.}
\label{fig:en_fluxes}
\end{figure}

For  ${B_0}=2$ and 10, the total energy flux is shown in Fig.~\ref{fig:total_flux}, while the individual fluxes (see Eq.~(\ref{partialflux})) are exhibited in Fig.~\ref{fig:en_fluxes}.  The plots are for a given snapshot during the evolved state.   Due to aforementioned reason (lack of averaging) and relatively smaller resolution, we do not observe constant energy fluxes.   

The most noticeable feature of the plots is the dominance of the inverse cascade of $\Pi^{u<}_{u>}(k_0)$ for $k < k_f$ when ${B_0}=10$.  This result is due to the quasi two-dimensionalization of the flow, and it is consistent with large kinetic energy at the large-scales near the equatorial region, discussed in the earlier section.  The other energy fluxes are several orders of magnitudes smaller than the maximum value of $\Pi^{u<}_{u>}(k_0)$.  

In addition to the inverse cascade of kinetic energy, we observe that for $k>k_f$, all the energy fluxes are positive, which is consistent with the earlier results by Debliquy {\em et al.}~\cite{Debliquy2005} for ${B_0}=0$.  Interestingly, $\Pi^{b<}_{b>} < 0$  for small wavenumbers ($k<k_{f}$) indicating inverse cascade of magnetic energy as well.  
  It is important to note however that  for $k > k_f$, $\Pi^{u<}_{u>}(k_0)$ is the most dominant flux and it is positive.    This is in contrast to the two-dimensional fluid turbulence in which the kinetic energy flux $\Pi^{u<}_{u>} \approx 0$ for $k>k_{f}$. The above feature is due to the forward energy transfer of $u_{\parallel}$.

\begin{figure}[h!]
\includegraphics[scale=1.0]{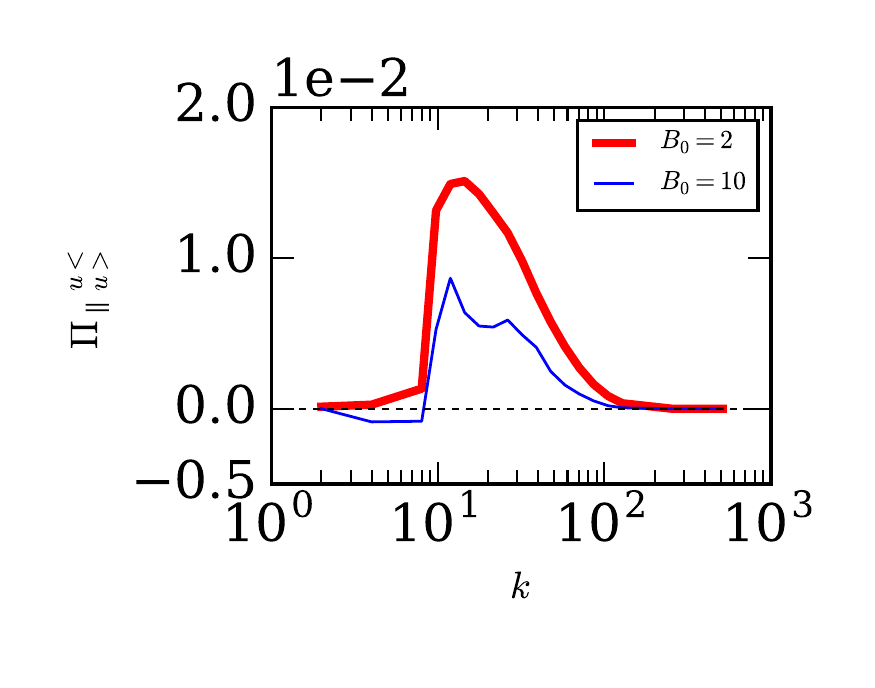}
\caption{Plot of the energy flux ${\Pi_{\parallel}^u} $ of the parallel component of the velocity field, $u_{\parallel}$.}
\label{fig:pllflux}
\end{figure}

For anisotropic flows, Reddy {\em et al.}~\cite{Reddy:POP2014} showed how to compute energy fluxes for the parallel and perpendicular components of the velocity fields. 
They showed that these fluxes are
\begin{eqnarray}
\Pi^u_\parallel & = & \sum_{k<k_0} \sum_{p>k_0} S_{\parallel}\left(\mathbf{k}\left|\mathbf{p}\right|\mathbf{q}\right)  \\
\Pi^u_\perp & = & \sum_{k<k_0} \sum_{p>k_0} S_{\perp}\left(\mathbf{k}\left|\mathbf{p}\right|\mathbf{q}\right) 
\end{eqnarray}
where
\begin{eqnarray}
S^u_{\parallel}\left(\mathbf{{k}\left|{p}\right|{q}}\right) & = & \Im\left\{ \left[\mathbf{k}\cdot\hat{\mathbf{u}}\left(\mathbf{q}\right)\right]\left[\hat{u}_{\parallel}^{\ast}\left(\mathbf{k}\right)\hat{u}_{\parallel}\left(\mathbf{p}\right)\right]\right\} \label{eq:S_pll} \\
S^u_{\perp}\left(\mathbf{k}\left|\mathbf{p}\right|\mathbf{q}\right) & = &
\Im\left\{ \left[\mathbf{k}\cdot\hat{\mathbf{u}}\left(\mathbf{q}\right)\right]\left[\mathbf{\hat{u}_{\perp}^{\ast}\left({k}\right)\cdot\hat{u}_{\perp}\left({p}\right)}\right]\right\}
\label{eq:S_perp}
\end{eqnarray}
where $\Im$ and $*$ stand for the imaginary and complex conjugate of the arguments.  Note that $ \Pi^{u<}_{u>} = \Pi^u_\parallel  + \Pi^u_\perp$.  It is easy to derive  the corresponding formulae for the magnetic energy by replacing $\hat{u}_{\parallel}$ and $\mathbf{\hat{u}_{\perp}}$ in Eqs.~(\ref{eq:S_pll}, \ref{eq:S_perp}) by $\hat{b}_{\parallel}$ and $\mathbf{\hat{b}_{\perp}}$ respectively.  In this paper, we report the above fluxes only for the velocity field since the magnetic energy is much smaller than the kinetic energy.  In Fig.~\ref{fig:pllflux} we plot $\Pi^u_\parallel$ that exhibits a  forward energy cascade of $u_{\parallel}$ at large wavenumbers.  The energy flux of the perpendicular component, $\Pi^u_\perp$ (not shown here), exhibits inverse cascade.  
The above observation is very similar to the quasi two-dimensional behaviour reported for quasi-static MHD turbulence by  Reddy {\em et al.}~\cite{Reddy:POP2014} and Favier {\em et al.}~\cite{Favier2010}--- $\mathbf{\hat{u}_{\perp}}$ exhibiting an inverse cascade at low wavenumbers, while $u_\parallel$ a forward cascade at large wavenumbers.  
We further note that kinetic helicity $H=\langle \bf u \cdot \nabla \times u\rangle $ in this quasi two-dimensional is a result of the correlation of the vertical velocity and the two dimensional vorticity $w_z=\partial_x u_y - \partial_y u_x$
thus the forward cascade of helicity is controlled by the forward cascade of the energy of the vertical component. 
The forward cascade of Helicity has been shown recently to alter the exponent of the energy spectrum  \cite{Sujovolsky2016}.

\begin{figure}[h!]
\hspace*{-2cm}
\includegraphics[scale=1.0, trim = 0cm 0cm 0cm 0cm, clip =true]{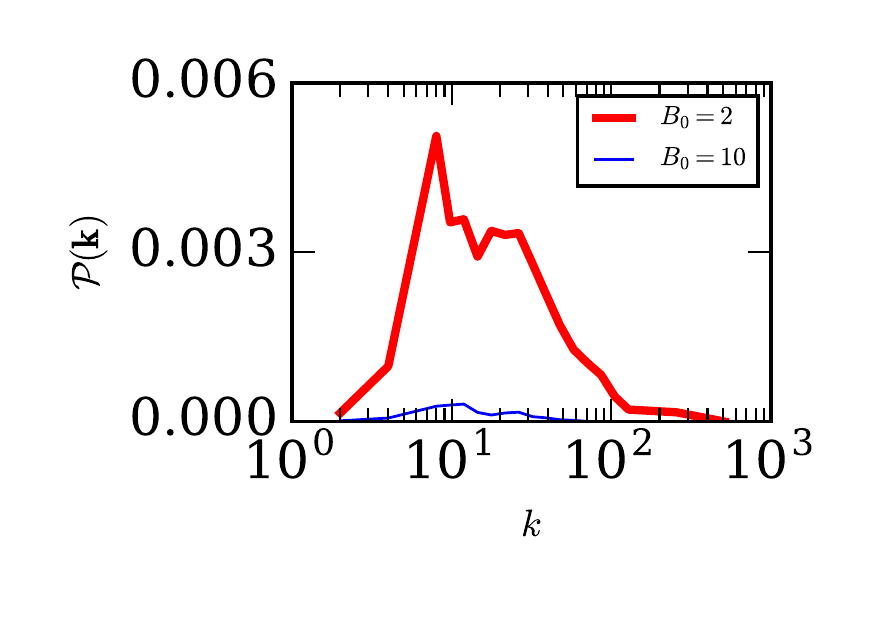} \\
\caption{Plot of $\mathcal{P}_\parallel({\bf k})$, the energy transfer rate from $u_{\perp}$ to $u_{\parallel}$ via pressure.}
\label{fig:pressureFlux}
\end{figure}

However $E^\perp_u$ and $E^\parallel_u$ are not independently conserved quantities.
$E^\perp_u$ energy can be transferred  to $E^\parallel_u$ and vice versa via pressure.
This transfer can be quantified by 
\begin{equation}
\mathcal{P}_\parallel({\bf k}) = \Im \left\{ [k_\parallel \hat{u}_\parallel({\bf k})]  P({\bf k}) \right\}
\end{equation}
as shown in \cite{Reddy:POP2014}.
A sum of the above over a wavenumber shell yields  energy transfer from ${\bf u}_\perp$ to $u_\parallel$  for that shell. 
The above energy transfer, plotted in Fig.~\ref{fig:pressureFlux}, reveals that this energy transfer is relatively weak 
for  ${B_0=10}$.  This feature may be due to relatively weak pressure and velocity fields.
 The energy transfer from ${\bf u}_\perp$ to $u_\parallel$ enhances $E^\parallel_u$, which is advected to larger wavenumbers.  Such features have been observed for quasi-static MHD~\cite{Reddy:POP2014}. The energy of the perpendicular component ($E^\perp_u$) however grows in the large scales in the presence of an inverse cascade.  This is not very significant for $B_0=2$ that has no inverse cascade, but it is dominant for $B_0=10$. 
Thus, $E^\parallel_u \sim E^\perp_u$ for $B_0=2$, but $E^\parallel_u \ll E^\perp_u$ for $B_0=10$ (see Table I).
As describe above and exhibited in Fig.~\ref{fig:pllflux}, $u_\parallel$ cascades forward to larger wavenumbers, 
which is the cause for the $A_{u}(k) = E_{u}^\perp (k)/(2E_{u}^\parallel(k)) < 1$ for large $k$.
We also observe that the energy transfers for the magnetic field may be coupled to the above transfers of the kinetic energy; this aspect needs to be investigated in detail.   

\begin{figure}[h!]
\hspace*{-1cm}
\vspace*{-2cm}
\includegraphics[scale=0.9, trim = 0cm 1cm 0cm 2cm, clip =true]{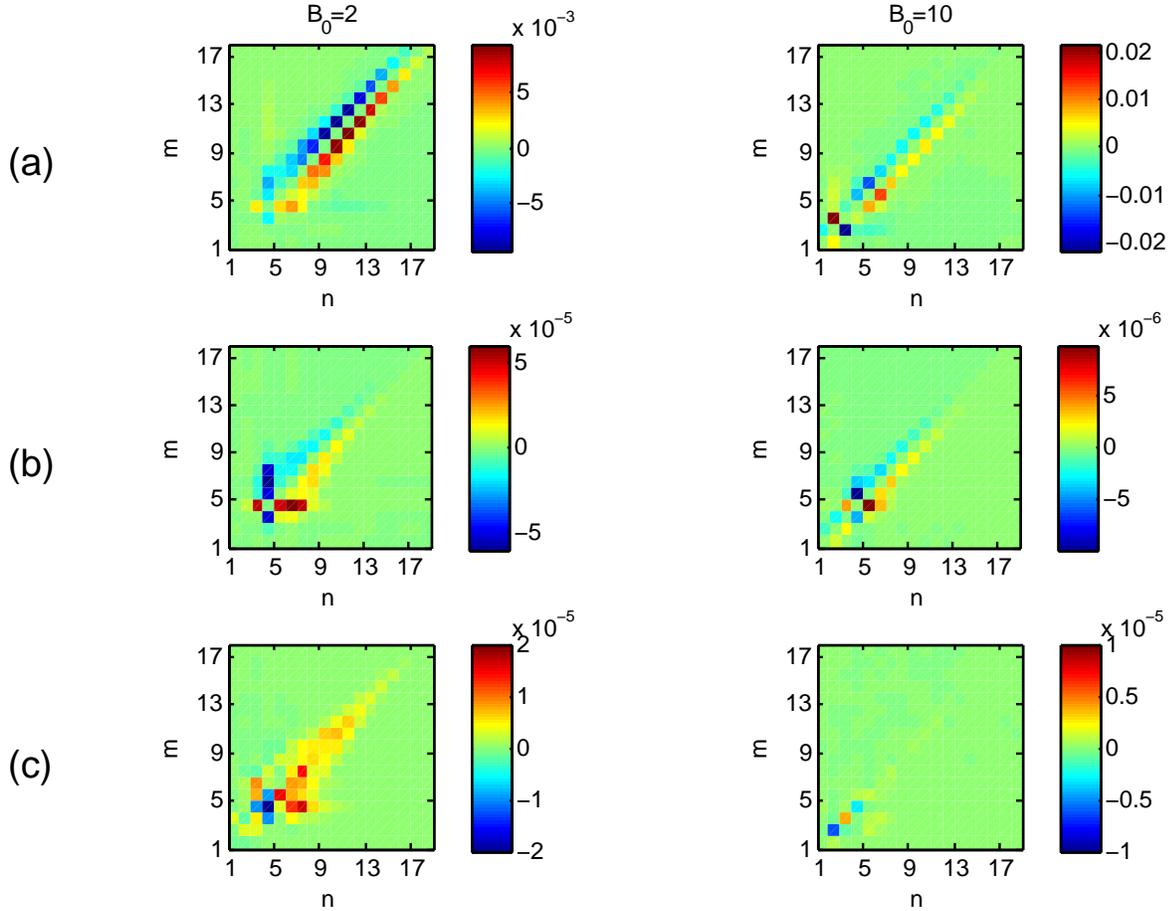} \\
\caption{Shell-to-shell energy transfers  (a) $U2U$,  (b) $B2B$, and  (c) $U2B$ for ${B_0=2}$(left column) and  
${B_0=10}$(right column). 
Here $m$ is the giver shell, and $n$ is the receiver shell.}
\label{fig:shell2shell}
\end{figure}

The energy flux describes the net energy emanating from a sphere.  More details on energy transfer is revealed by the shell-to-shell energy transfer rates.  For fluid turbulence, we have shell-to-shell transfers for the velocity field.  However, for MHD turbulence we have velocity-to-velocity ($U2U$), magnetic-to-magnetic ($B2B$), and kinetic-to-magnetic ($U2B$) shell-to-shell energy transfers~\cite{Dar2001,Alexakis2005,Debliquy2005, Teaca:PRE2009}.   The   energy transfer from wavenumber shell $m$ of field $X$ to wavenumber shell $n$ of field $Y$ is defined as ($X,Y$ are either velocity or magnetic field):
\begin{eqnarray}
\mathcal{T}^{u,u}_{n,m} & = &  \sum_{k\in \mathcal{D}_n} \sum_{p \in \mathcal{D}_m} \Im\{[{\bf k} \cdot {\bf \hat{u}({\bf q})} ][{\bf \hat{u}}({\bf p}) \cdot {\bf  \hat{u}}^{*} ({\bf k})]\} 
=-\int {[\bf u_k (u\cdot\nabla)] u_p} dx^3\\
\mathcal{T}^{b,b}_{n,m} & = &  \sum_{k\in \mathcal{D}_n} \sum_{p \in \mathcal{D}_m} \Im\{[{\bf k} \cdot {\bf \hat{u}({\bf q})} ][{\bf \hat{b}}({\bf p}) \cdot {\bf  \hat{b}}^{*} ({\bf k})]\}
=-\int {[\bf b_k (u\cdot\nabla)] b_p} dx^3\\
\mathcal{T}^{b,u}_{n,m} & = & - \sum_{k\in \mathcal{D}_n} \sum_{p \in \mathcal{D}_m} \Im\{[{\bf k} \cdot {\bf \hat{u}({\bf q})} ][{\bf \hat{u}}({\bf p}) \cdot {\bf  \hat{b}}^{*} ({\bf k})]\} 
=+\int {[\bf b_k (b\cdot\nabla)] u_p} dx^3
\end{eqnarray}
 For the shell-to-shell energy transfers we divide the wavenumber space into 19 concentric shells with their centres at k = (0, 0, 0).  The inner and outer radii of the $n$th shell are $k_{n-1}$ and $k_n$ respectively, where $k_n= 0$, 2.0, 4.0, 8.0, 9.8, 12.0, 14.8, 18.1, 22.2, 27.2, 33.4, 40.9, 50.2, 61.5, 75.4, 92.5, 113.4, 139.0, 170.5, and 341.0.  The aforementioned radii are chosen using the same algorithm as those used for the computing the radii of the spheres for the flux computations.
  In Fig.~\ref{fig:shell2shell}, we present the shell-to-shell energy transfer rates, $T^{uu}_{nm}$,   $T^{bb}_{nm}$, and $T^{bu}_{nm}$ for $B_0=2$ (left column) and $B_0=10$ ( right column).  

The $U2U$ and $B2B$ transfers for ${B_0} = 2$, exhibited in Fig.~\ref{fig:shell2shell}(a) is similar to those reported by Alexakis {\em et al.}~\cite{Alexakis2005}, Debliquy {\em et al.}~\cite{Debliquy2005},  and ~Carati {\em et al.}~\cite{Carati2006}  for ${B_0}=0$ forward and local $U2U$ and $B2B$ transfers, that is, the most energy transfers are from shell $m-1$ to shell $m$.  The $U2B$ transfer is from  shell $m$ of the velocity field to  shell $m$ of the magnetic field, which is because the velocity field dominates the magnetic field~\cite{Debliquy2005}; this feature is exactly opposite to that for  ${B_0}=0$~\cite{Alexakis2005,Debliquy2005,Carati2006}  because $E_b > E_u$ for the ${B_0}=0$ case.

For ${B_0 =10}$ (see Fig.~\ref{fig:shell2shell}), $U2U$is the most dominant transfer, and the $U2U$ and $B2B$ shell-to-shell transfer exhibits 
inverse energy transfers for the 3rd and 4th shell ($k<k_f$), i.e., from the 4th shell to the 3rd shell.  This result is consistent with the 
inverse cascades of kinetic and magnetic energies for $k<k_f$ (see Fig.~\ref{fig:en_fluxes}).  The $U2B$ transfers are  nonzero only for $k<k_f$.

\section{Summary and Discussion} 
In this paper we analyzed the anisotropy induced by a constant magnetic field in MHD turbulence.  Here we provide semiquantitative   picture of the above phenomena.  Shear Alfv\'{e}n  modes are linear excitations of MHD flows, and they are governed by equations:
\begin{equation}
\frac{d \hat{\bf u}({\bf k})}{dt} = i ({\bf B_0 \cdot k})  \hat{\bf b}({\bf k});~~~\frac{d \hat{\bf b}({\bf k})}{dt} = i ({\bf B_0 \cdot k})  \hat{\bf u}({\bf k})
\label{eq:Alfven}
\end{equation}
The above equations have valid wave solutions when ${\bf B_0 \cdot k} \ne 0$, that is, for wave vectors off from the plane perpendicular to the mean magnetic field.  For such modes, in Eq.~(\ref{eq:uk},\ref{eq:bk}), $({\bf B}_0 \cdot {\bf k}) {\bf u(k)}$ and $({\bf B}_0 \cdot {\bf k}) {\bf b(k)}$ dominates the nonlinear term.  Earlier, Galtier {\em et al.}~\cite{Galtier2000} had analysed the weak turbulence limit of MHD turbulence for large ${B_0}$ and showed that $E(k_\perp) \sim k_\perp^{-2}$.  

 For the Fourier modes with $k_\parallel = 0$, the linear terms dropout of Eqs.~(\ref{eq:uk},\ref{eq:bk}) 
and the nonlinear terms dominate the flow with dynamics.  In addition, for large $B_0$, $b^2 \ll u^2$ (see Table I).  Since $k_\parallel = 0$ for such modes, the modes have interactions  similar to two-dimensional hydrodynamic turbulence.  These interactions lead to two-dimensionalization of the flow. The reason for $b^2 \ll u^2$ is unclear at present, but it may be due to the absence of share Alfv\'{e}n waves for modes with $k_\parallel =0$.    To sum up, for the Fourier modes with $k_\parallel \ne 0$, we obtain Alfv\'{e}nic fluctuations, which are described by Eqs.~(\ref{eq:Alfven}) in the linear limit.  However, for large $B_0$, the fluctuations corresponding to these modes are weak compared to the vortical structures.  Thus the flow is dominated by the $k_\parallel = 0$ modes.  These arguments provide qualitative picture for the emergence of quasi two-dimensional vortices in MHD turbulence with strong $B_0$.  The above behaviour has strong  similarities with the vortical structures observed in rotating and quasi-static MHD turbulence~\cite{Reddy:POF2014}.

The dominance of these modes leads then to an anisotropic distribution of the velocity components with the perpendicular components 
dominating in the large scales due to the inverse cascade of $E_\perp$ while the parallel components dominate in the small scales due to the forward cascade of $E_\parallel$.  This leads to the formation of the observed vortical structures.   

In  summary, we show how strong mean magnetic field makes the MHD turbulence quasi two-dimensional.
This conclusion is borne out in the global-energy anisotropy parameter, ring spectrum, energy flux, and shell-to-shell energy transfers.  
The flow has strong similarities with those observed in rotating and quasi-static MHD turbulence.  
Detailed dynamical connections between these flows need to be explored in a future work.

\section{Acknowledgments}
We thank Sandeep Reddy, Abhishek Kumar, Biplab Dutta and Rohit Kumar for valuable discussions.
Our  numerical  simulations  were  performed at HPC2013 and Chaos clusters of IIT Kanpur.
This work was supported by the research grants 4904-A from Indo-French Centre for the Promotion of Advanced Research (IFCPAR/CEFIPRA), 
SERB/F/3279/2013-14 from the Science and Engineering Research Board, India, and  Project A9 via SFB-TR24 from DFG Germany.


\begin{thebibliography}{36}
\expandafter\ifx\csname natexlab\endcsname\relax\def\natexlab#1{#1}\fi
\expandafter\ifx\csname bibnamefont\endcsname\relax
  \def\bibnamefont#1{#1}\fi
\expandafter\ifx\csname bibfnamefont\endcsname\relax
  \def\bibfnamefont#1{#1}\fi
\expandafter\ifx\csname citenamefont\endcsname\relax
  \def\citenamefont#1{#1}\fi
\expandafter\ifx\csname url\endcsname\relax
  \def\url#1{\texttt{#1}}\fi
\expandafter\ifx\csname urlprefix\endcsname\relax\def\urlprefix{URL }\fi
\providecommand{\bibinfo}[2]{#2}
\providecommand{\eprint}[2][]{\url{#2}}

\bibitem[{\citenamefont{Alexakis}(2013)}]{Alexakis2013}
\bibinfo{author}{\bibfnamefont{A.}~\bibnamefont{Alexakis}},
  \bibinfo{journal}{Phys. Rev. Lett.} \textbf{\bibinfo{volume}{110}},
  \bibinfo{pages}{084502} (\bibinfo{year}{2013}).

\bibitem[{\citenamefont{{Iroshnikov}}(1963)}]{Iroshnikov1963}
\bibinfo{author}{\bibfnamefont{P.~S.} \bibnamefont{{Iroshnikov}}},
  \bibinfo{journal}{Soviet Astronomy} \textbf{\bibinfo{volume}{40}},
  \bibinfo{pages}{742} (\bibinfo{year}{1963}).

\bibitem[{\citenamefont{{Kraichnan}}(1965)}]{Kraichnan1965}
\bibinfo{author}{\bibfnamefont{R.~H.} \bibnamefont{{Kraichnan}}},
  \bibinfo{journal}{Physics of Fluids} \textbf{\bibinfo{volume}{8}},
  \bibinfo{pages}{1385} (\bibinfo{year}{1965}).

\bibitem[{\citenamefont{Shebalin et~al.}(1983)\citenamefont{Shebalin,
  Matthaeus, and Montgomery}}]{Shebalin1983}
\bibinfo{author}{\bibfnamefont{J.~V.} \bibnamefont{Shebalin}},
  \bibinfo{author}{\bibfnamefont{W.~H.} \bibnamefont{Matthaeus}},
  \bibnamefont{and}
  \bibinfo{author}{\bibfnamefont{D.}~\bibnamefont{Montgomery}},
  \bibinfo{journal}{J. Plasma Physics} \textbf{\bibinfo{volume}{29}},
  \bibinfo{pages}{525} (\bibinfo{year}{1983}).

\bibitem[{\citenamefont{Zank and Matthaeus}(1993)}]{Zank1993}
\bibinfo{author}{\bibfnamefont{G.~P.} \bibnamefont{Zank}} \bibnamefont{and}
  \bibinfo{author}{\bibfnamefont{W.}~\bibnamefont{Matthaeus}},
  \bibinfo{journal}{Physics of Fluids A: Fluid Dynamics (1989-1993)}
  \textbf{\bibinfo{volume}{5}}, \bibinfo{pages}{257} (\bibinfo{year}{1993}).

\bibitem[{\citenamefont{Oughton et~al.}(1994)\citenamefont{Oughton, Priest, and
  Matthaeus}}]{Oughton1994}
\bibinfo{author}{\bibfnamefont{S.}~\bibnamefont{Oughton}},
  \bibinfo{author}{\bibfnamefont{E.~R.} \bibnamefont{Priest}},
  \bibnamefont{and} \bibinfo{author}{\bibfnamefont{W.~H.}
  \bibnamefont{Matthaeus}}, \bibinfo{journal}{Journal of Fluid Mechanics}
  \textbf{\bibinfo{volume}{280}}, \bibinfo{pages}{95} (\bibinfo{year}{1994}).

\bibitem[{\citenamefont{{Verma}}(1999)}]{Verma:PP1999}
\bibinfo{author}{\bibfnamefont{M.~K.} \bibnamefont{{Verma}}},
  \bibinfo{journal}{Phys. Plasmas} \textbf{\bibinfo{volume}{6}},
  \bibinfo{pages}{1455} (\bibinfo{year}{1999}).

\bibitem[{\citenamefont{Verma}(2004)}]{Verma2004}
\bibinfo{author}{\bibfnamefont{M.~K.} \bibnamefont{Verma}},
  \bibinfo{journal}{Physics reports} \textbf{\bibinfo{volume}{401}},
  \bibinfo{pages}{229} (\bibinfo{year}{2004}).

\bibitem[{\citenamefont{{Verma}}(2001)}]{Verma:PRE2001}
\bibinfo{author}{\bibfnamefont{M.}~\bibnamefont{{Verma}}},
  \bibinfo{journal}{Phys. Rev. E} \textbf{\bibinfo{volume}{64}},
  \bibinfo{pages}{26305} (\bibinfo{year}{2001}).

\bibitem[{\citenamefont{Goldreich and Sridhar}(1995)}]{Goldreich1995}
\bibinfo{author}{\bibfnamefont{P.}~\bibnamefont{Goldreich}} \bibnamefont{and}
  \bibinfo{author}{\bibfnamefont{S.}~\bibnamefont{Sridhar}},
  \bibinfo{journal}{The Astrophysical Journal} \textbf{\bibinfo{volume}{438}},
  \bibinfo{pages}{763} (\bibinfo{year}{1995}).

\bibitem[{\citenamefont{Boldyrev}(2006)}]{Boldyrev2006}
\bibinfo{author}{\bibfnamefont{S.}~\bibnamefont{Boldyrev}},
  \bibinfo{journal}{Phys. Rev. Lett.} \textbf{\bibinfo{volume}{96}},
  \bibinfo{pages}{115002} (\bibinfo{year}{2006}).

\bibitem[{\citenamefont{{Boldyrev} and {Perez}}(2009)}]{Boldyrev:PhRv2009}
\bibinfo{author}{\bibfnamefont{S.}~\bibnamefont{{Boldyrev}}} \bibnamefont{and}
  \bibinfo{author}{\bibfnamefont{J.~C.} \bibnamefont{{Perez}}},
  \bibinfo{journal}{Physical Review Letters} \textbf{\bibinfo{volume}{103}},
  \bibinfo{pages}{225001} (\bibinfo{year}{2009}).

\bibitem[{\citenamefont{Perez et~al.}(2012)\citenamefont{Perez, Mason,
  Boldyrev, and Cattaneo}}]{Boldyrev2012}
\bibinfo{author}{\bibfnamefont{J.~C.} \bibnamefont{Perez}},
  \bibinfo{author}{\bibfnamefont{J.}~\bibnamefont{Mason}},
  \bibinfo{author}{\bibfnamefont{S.}~\bibnamefont{Boldyrev}}, \bibnamefont{and}
  \bibinfo{author}{\bibfnamefont{F.}~\bibnamefont{Cattaneo}},
  \bibinfo{journal}{Phys. Rev. X} \textbf{\bibinfo{volume}{2}},
  \bibinfo{pages}{041005} (\bibinfo{year}{2012}).

\bibitem[{\citenamefont{{Perez} et~al.}(2014)\citenamefont{{Perez}, {Mason},
  {Boldyrev}, and {Cattaneo}}}]{Boldyrev:ApJ2014}
\bibinfo{author}{\bibfnamefont{J.~C.} \bibnamefont{{Perez}}},
  \bibinfo{author}{\bibfnamefont{J.}~\bibnamefont{{Mason}}},
  \bibinfo{author}{\bibfnamefont{S.}~\bibnamefont{{Boldyrev}}},
  \bibnamefont{and}
  \bibinfo{author}{\bibfnamefont{F.}~\bibnamefont{{Cattaneo}}},
  \bibinfo{journal}{Astrophys. J. Lett.} \textbf{\bibinfo{volume}{793}},
  \bibinfo{pages}{L13} (\bibinfo{year}{2014}).

\bibitem[{\citenamefont{Beresnyak and A.}(2009)}]{Beresnyak:ApJ2009}
\bibinfo{author}{\bibfnamefont{A.}~\bibnamefont{Beresnyak}} \bibnamefont{and}
  \bibinfo{author}{\bibfnamefont{L.}~\bibnamefont{A.}},
  \bibinfo{journal}{Astrophys. J} \textbf{\bibinfo{volume}{702}},
  \bibinfo{pages}{1190} (\bibinfo{year}{2009}).

\bibitem[{\citenamefont{{Beresnyak}}(2011)}]{Beresnyak:PhRvL2011}
\bibinfo{author}{\bibfnamefont{A.}~\bibnamefont{{Beresnyak}}},
  \bibinfo{journal}{Physical Review Letters} \textbf{\bibinfo{volume}{106}},
  \bibinfo{pages}{075001} (\bibinfo{year}{2011}).

\bibitem[{\citenamefont{{Beresnyak}}(2014)}]{Beresnyak:ApJ2014}
\bibinfo{author}{\bibfnamefont{A.}~\bibnamefont{{Beresnyak}}},
  \bibinfo{journal}{Astrophys. J. Lett.} \textbf{\bibinfo{volume}{784}},
  \bibinfo{pages}{L20} (\bibinfo{year}{2014}).

\bibitem[{\citenamefont{{Galtier} et~al.}(2000)\citenamefont{{Galtier},
  {Nazarenko}, {Newell}, and {Pouquet}}}]{Galtier2000}
\bibinfo{author}{\bibfnamefont{S.}~\bibnamefont{{Galtier}}},
  \bibinfo{author}{\bibfnamefont{S.~V.} \bibnamefont{{Nazarenko}}},
  \bibinfo{author}{\bibfnamefont{A.~C.} \bibnamefont{{Newell}}},
  \bibnamefont{and}
  \bibinfo{author}{\bibfnamefont{A.}~\bibnamefont{{Pouquet}}},
  \bibinfo{journal}{J. of Plasma Physics} \textbf{\bibinfo{volume}{63}},
  \bibinfo{pages}{447} (\bibinfo{year}{2000}).

\bibitem[{\citenamefont{Alexakis}(2011)}]{Alexakis2011}
\bibinfo{author}{\bibfnamefont{A.}~\bibnamefont{Alexakis}},
  \bibinfo{journal}{Phys. Rev. E} \textbf{\bibinfo{volume}{84}},
  \bibinfo{pages}{056330} (\bibinfo{year}{2011}).

\bibitem[{\citenamefont{{Reddy} and {Verma}}(2014)}]{Reddy:POF2014}
\bibinfo{author}{\bibfnamefont{K.~S.} \bibnamefont{{Reddy}}} \bibnamefont{and}
  \bibinfo{author}{\bibfnamefont{M.~K.} \bibnamefont{{Verma}}},
  \bibinfo{journal}{Physics of Fluids} \textbf{\bibinfo{volume}{26}},
  \bibinfo{eid}{025109} (\bibinfo{year}{2014}).

\bibitem[{\citenamefont{Reddy et~al.}(2014)\citenamefont{Reddy, Kumar, and
  Verma}}]{Reddy:POP2014}
\bibinfo{author}{\bibfnamefont{K.~S.} \bibnamefont{Reddy}},
  \bibinfo{author}{\bibfnamefont{R.}~\bibnamefont{Kumar}}, \bibnamefont{and}
  \bibinfo{author}{\bibfnamefont{M.~K.} \bibnamefont{Verma}},
  \bibinfo{journal}{Physics of Plasmas} \textbf{\bibinfo{volume}{21}},
  \bibinfo{pages}{102310} (\bibinfo{year}{2014}).

\bibitem[{\citenamefont{{Gallet} and {Doering}}(2015)}]{Gallet2015}
\bibinfo{author}{\bibfnamefont{B.}~\bibnamefont{{Gallet}}} \bibnamefont{and}
  \bibinfo{author}{\bibfnamefont{C.~R.} \bibnamefont{{Doering}}},
  \bibinfo{journal}{Journal of Fluid Mechanics} \textbf{\bibinfo{volume}{773}},
  \bibinfo{pages}{154} (\bibinfo{year}{2015}).

\bibitem[{\citenamefont{Alexakis et~al.}(2007)\citenamefont{Alexakis, Bigot,
  Politano, and Galtier}}]{Alexakis2007}
\bibinfo{author}{\bibfnamefont{A.}~\bibnamefont{Alexakis}},
  \bibinfo{author}{\bibfnamefont{B.}~\bibnamefont{Bigot}},
  \bibinfo{author}{\bibfnamefont{H.}~\bibnamefont{Politano}}, \bibnamefont{and}
  \bibinfo{author}{\bibfnamefont{S.}~\bibnamefont{Galtier}},
  \bibinfo{journal}{Phys. Rev. E} \textbf{\bibinfo{volume}{76}},
  \bibinfo{pages}{056313} (\bibinfo{year}{2007}).

\bibitem[{\citenamefont{Teaca et~al.}(2009)\citenamefont{Teaca, Verma, Knaepen,
  and Carati}}]{Teaca:PRE2009}
\bibinfo{author}{\bibfnamefont{B.}~\bibnamefont{Teaca}},
  \bibinfo{author}{\bibfnamefont{M.~K.} \bibnamefont{Verma}},
  \bibinfo{author}{\bibfnamefont{B.}~\bibnamefont{Knaepen}}, \bibnamefont{and}
  \bibinfo{author}{\bibfnamefont{D.}~\bibnamefont{Carati}},
  \bibinfo{journal}{Phys. Rev. E} \textbf{\bibinfo{volume}{79}},
  \bibinfo{pages}{046312} (\bibinfo{year}{2009}).

\bibitem[{\citenamefont{Roberts}(1967)}]{Roberts1967}
\bibinfo{author}{\bibfnamefont{P.~H.} \bibnamefont{Roberts}},
  \emph{\bibinfo{title}{{An Introduction to Magnetohydrodynamics}}}
  (\bibinfo{publisher}{New York: Elsevier}, \bibinfo{year}{1967}).

\bibitem[{\citenamefont{Mininni et~al.}(2011)\citenamefont{Mininni, Rosenberg,
  Reddy, and Pouquet}}]{ghost}
\bibinfo{author}{\bibfnamefont{P.~D.} \bibnamefont{Mininni}},
  \bibinfo{author}{\bibfnamefont{D.}~\bibnamefont{Rosenberg}},
  \bibinfo{author}{\bibfnamefont{R.}~\bibnamefont{Reddy}}, \bibnamefont{and}
  \bibinfo{author}{\bibfnamefont{A.}~\bibnamefont{Pouquet}},
  \bibinfo{journal}{Parallel Computing} \textbf{\bibinfo{volume}{37}},
  \bibinfo{pages}{316 } (\bibinfo{year}{2011}).

\bibitem[{\citenamefont{Verma et~al.}(2013)\citenamefont{Verma, Chatterjee,
  Reddy, Yadav, Paul, Chandra, and Samtaney}}]{tarang}
\bibinfo{author}{\bibfnamefont{M.~K.} \bibnamefont{Verma}},
  \bibinfo{author}{\bibfnamefont{A.}~\bibnamefont{Chatterjee}},
  \bibinfo{author}{\bibfnamefont{K.~S.} \bibnamefont{Reddy}},
  \bibinfo{author}{\bibfnamefont{R.~K.} \bibnamefont{Yadav}},
  \bibinfo{author}{\bibfnamefont{S.}~\bibnamefont{Paul}},
  \bibinfo{author}{\bibfnamefont{M.}~\bibnamefont{Chandra}}, \bibnamefont{and}
  \bibinfo{author}{\bibfnamefont{R.}~\bibnamefont{Samtaney}},
  \bibinfo{journal}{Pramana} \textbf{\bibinfo{volume}{81}},
  \bibinfo{pages}{617} (\bibinfo{year}{2013}).

\bibitem[{\citenamefont{{Favier} et~al.}(2010)\citenamefont{{Favier},
  {Godeferd}, {Cambon}, and {Delache}}}]{Favier2010}
\bibinfo{author}{\bibfnamefont{B.}~\bibnamefont{{Favier}}},
  \bibinfo{author}{\bibfnamefont{F.~S.} \bibnamefont{{Godeferd}}},
  \bibinfo{author}{\bibfnamefont{C.}~\bibnamefont{{Cambon}}}, \bibnamefont{and}
  \bibinfo{author}{\bibfnamefont{A.}~\bibnamefont{{Delache}}},
  \bibinfo{journal}{Physics of Fluids} \textbf{\bibinfo{volume}{22}},
  \bibinfo{eid}{075104} (\bibinfo{year}{2010}).

\bibitem[{\citenamefont{TenBarge et~al.}(2012)\citenamefont{TenBarge, Podesta,
  Klein, and Howes}}]{0004-637X-753-2-107}
\bibinfo{author}{\bibfnamefont{J.~M.} \bibnamefont{TenBarge}},
  \bibinfo{author}{\bibfnamefont{J.~J.} \bibnamefont{Podesta}},
  \bibinfo{author}{\bibfnamefont{K.~G.} \bibnamefont{Klein}}, \bibnamefont{and}
  \bibinfo{author}{\bibfnamefont{G.~G.} \bibnamefont{Howes}},
  \bibinfo{journal}{The Astrophysical Journal} \textbf{\bibinfo{volume}{753}},
  \bibinfo{pages}{107} (\bibinfo{year}{2012}).

\bibitem[{\citenamefont{Alexandrova et~al.}(2008)\citenamefont{Alexandrova,
  Carbone, Veltri, and Sorriso-Valvo}}]{0004-637X-674-2-1153}
\bibinfo{author}{\bibfnamefont{O.}~\bibnamefont{Alexandrova}},
  \bibinfo{author}{\bibfnamefont{V.}~\bibnamefont{Carbone}},
  \bibinfo{author}{\bibfnamefont{P.}~\bibnamefont{Veltri}}, \bibnamefont{and}
  \bibinfo{author}{\bibfnamefont{L.}~\bibnamefont{Sorriso-Valvo}},
  \bibinfo{journal}{The Astrophysical Journal} \textbf{\bibinfo{volume}{674}},
  \bibinfo{pages}{1153} (\bibinfo{year}{2008}).

\bibitem[{\citenamefont{Dar et~al.}(2001)\citenamefont{Dar, Verma, and
  Eswaran}}]{Dar2001}
\bibinfo{author}{\bibfnamefont{G.}~\bibnamefont{Dar}},
  \bibinfo{author}{\bibfnamefont{M.~K.} \bibnamefont{Verma}}, \bibnamefont{and}
  \bibinfo{author}{\bibfnamefont{V.}~\bibnamefont{Eswaran}},
  \bibinfo{journal}{Physica D: Nonlinear Phenomena}
  \textbf{\bibinfo{volume}{157}}, \bibinfo{pages}{207} (\bibinfo{year}{2001}).

\bibitem[{\citenamefont{{Alexakis} et~al.}(2005)\citenamefont{{Alexakis},
  {Mininni}, and {Pouquet}}}]{Alexakis2005}
\bibinfo{author}{\bibfnamefont{A.}~\bibnamefont{{Alexakis}}},
  \bibinfo{author}{\bibfnamefont{P.~D.} \bibnamefont{{Mininni}}},
  \bibnamefont{and}
  \bibinfo{author}{\bibfnamefont{A.}~\bibnamefont{{Pouquet}}},
  \bibinfo{journal}{Phys. Rev. E} \textbf{\bibinfo{volume}{72}},
  \bibinfo{eid}{046301} (\bibinfo{year}{2005}).

\bibitem[{\citenamefont{{Debliquy} et~al.}(2005)\citenamefont{{Debliquy},
  {Verma}, and {Carati}}}]{Debliquy2005}
\bibinfo{author}{\bibfnamefont{O.}~\bibnamefont{{Debliquy}}},
  \bibinfo{author}{\bibfnamefont{M.~K.} \bibnamefont{{Verma}}},
  \bibnamefont{and} \bibinfo{author}{\bibfnamefont{D.}~\bibnamefont{{Carati}}},
  \bibinfo{journal}{Physics of Plasmas} \textbf{\bibinfo{volume}{12}},
  \bibinfo{eid}{042309} (\bibinfo{year}{2005}).

\bibitem[{\citenamefont{Mininni et~al.}(2006)\citenamefont{Mininni, Pouquet,
  and Montgomery}}]{MininniPRE2005}
\bibinfo{author}{\bibfnamefont{P.~D.} \bibnamefont{Mininni}},
  \bibinfo{author}{\bibfnamefont{A.~G.} \bibnamefont{Pouquet}},
  \bibnamefont{and} \bibinfo{author}{\bibfnamefont{D.~C.}
  \bibnamefont{Montgomery}}, \bibinfo{journal}{Phys. Rev. Lett.}
  \textbf{\bibinfo{volume}{97}}, \bibinfo{pages}{244503}
  (\bibinfo{year}{2006}).

\bibitem[{\citenamefont{{Sujovolsky} and {Mininni}}(2016)}]{Sujovolsky2016}
\bibinfo{author}{\bibfnamefont{N.~E.} \bibnamefont{{Sujovolsky}}}
  \bibnamefont{and} \bibinfo{author}{\bibfnamefont{P.~D.}
  \bibnamefont{{Mininni}}}, \bibinfo{journal}{ArXiv e-prints}
  (\bibinfo{year}{2016}), \eprint{1606.04026}.

\bibitem[{\citenamefont{Carati et~al.}(2006)\citenamefont{Carati, Debliquy,
  Knaepen, Teaca, and Verma}}]{Carati2006}
\bibinfo{author}{\bibfnamefont{D.}~\bibnamefont{Carati}},
  \bibinfo{author}{\bibfnamefont{O.}~\bibnamefont{Debliquy}},
  \bibinfo{author}{\bibfnamefont{B.}~\bibnamefont{Knaepen}},
  \bibinfo{author}{\bibfnamefont{B.}~\bibnamefont{Teaca}}, \bibnamefont{and}
  \bibinfo{author}{\bibfnamefont{M.~K.} \bibnamefont{Verma}},
  \bibinfo{journal}{Journal of Turbulence} \textbf{\bibinfo{volume}{7}},
  \bibinfo{pages}{N51} (\bibinfo{year}{2006}).

\end{thebibliography}

\end{document}